\RequirePackage{fix-cm}
\documentclass[twocolumn]{svjour3}       
\smartqed  

\usepackage{makeidx}  
\usepackage[hyphens]{url}

\usepackage{easyReview}
\usepackage{ifthen}
\usepackage{amssymb}
\usepackage{cite}
\usepackage{color}
\usepackage{colortbl}
\usepackage{graphicx}
\usepackage{amsmath}
\usepackage{tabularx}
\usepackage{comment}
\usepackage{booktabs}
\usepackage{multirow}
\usepackage{hhline}
\usepackage{soul}
\usepackage{lineno,hyperref}
\usepackage{wrapfig}
\usepackage{tikz}
\usepackage{multirow}
\usepackage{framed}
\usepackage{subcaption}
\usetikzlibrary{positioning}

\usepackage{algpseudocode,algorithm,algorithmicx}

\algrenewcommand\algorithmicrequire{\textbf{input:}}
\algrenewcommand\algorithmicindent{1em}%

\makeatletter
\let\OldStatex\Statex
\renewcommand{\Statex}[1][3]{%
  \setlength\@tempdima{\algorithmicindent}%
  \OldStatex\hskip\dimexpr#1\@tempdima\relax}
\makeatother

\newcommand*\circled[1]{\tikz[baseline=(char.base)]{
            \node[shape=circle,draw,inner sep=1.2pt] (char) {#1};}}

\newcommand{\eg}{e.g.,~}							
\newcommand{\ie}{i.e.,~}							
\newcommand{\etal}{~\textit{et al.}}					
\newcommand{\Fig}[1]{Fig.~\ref{#1}}  			
\newcommand{\Table}[1]{Table~\ref{#1}}	    
\newcommand{\Sect}[1]{Section~\ref{#1}}	  
\newcommand{\Code}[1]{\texttt{\small{#1}}}	

\begin{document}
\title{Recommending Metamodel Concepts during Modeling Activities with Pre-Trained Language Models
}


\author{Martin Weyssow \and Houari Sahraoui \and Eugene Syriani}


\institute{M. Weyssow \at
              DIRO, Universit\'e de Montr\'eal, Montreal, Canada \\
              \email{martin.weyssow@umontreal.ca}
           \and
           H. Sahraoui \at
              DIRO, Universit\'e de Montr\'eal, Montreal, Canada \\
              \email{sahraouh@iro.umontreal.ca}
           \and
           E. Syriani \at
              DIRO, Universit\'e de Montr\'eal, Montreal, Canada \\
              \email{syriani@iro.umontreal.ca}
}

\date{Received: 2 April 2021 / Accepted: 15 December 2021}

\maketitle

\begin{abstract}
The design of conceptually sound metamodels that embody proper semantics in relation to the application domain is particularly tedious in Model-Driven Engineering. As metamodels define complex relationships between domain concepts, it is crucial for a modeler to define these concepts thoroughly while being consistent with respect to the application domain.
We propose an approach to assist a modeler in the design of metamodel by recommending relevant domain concepts in several modeling scenarios. Our approach does not require knowledge from the domain or to hand-design completion rules. Instead, we design a fully data-driven approach using a deep learning model that is able to abstract domain concepts by learning from both structural and lexical metamodel properties in a corpus of thousands of independent metamodels. 
We evaluate our approach on a test set containing 166 metamodels, unseen during the model training, with more than 5000 test samples. Our preliminary results show that the trained model is able to provide accurate top-$5$ lists of relevant recommendations for concept renaming scenarios. Although promising, the results are less compelling for the scenario of the iterative construction of the metamodel, in part because of the conservative strategy we use to evaluate the recommendations.

\end{abstract}

\section{Introduction}
\label{sec:introduction}

\textit{Model-Driven Engineering} (MDE) is an increasingly popular software development methodology that has proven to be valuable in real-world environments and for the development of large-scale systems~\cite{baker2005model, mohagheghi2013empirical}. In 2014, Whittle\etal{} found in their study that the use of MDE in practice was even more common than the MDE community thought~\cite{usage_mde_practice}. With the ever-increasing complexity of software systems, MDE provides a framework of practice that facilitates the development and maintenance of such complex systems. Nevertheless, practitioners may be reluctant to the use of such technology, which requires considerable human efforts during the early stage of the system's development~\cite{ATKINSON2007105}. A recent work highlighted the need for \textit{Intelligent Modeling Assistants} (IMA) to better support the development of MDE-based systems~\cite{mussbacher2020opportunities}. Given the aforementioned difficulties in MDE in conjunction with the tremendous amount of available data, these previous works have prompted researchers to focus on the development of data-driven recommender systems. Therefore, the design of efficient IMAs remains one of the key challenges in MDE to ease the cognitive load of modelers and spread the discipline further by reducing the effort and cost associated to its application. 

One promising way to assist modelers is to recommend proper names to software elements that are consistent with the considered application domain. 
A previous study showed that a consistent naming of elements plays a critical role in the quality of the metamodels~\cite{lopez2014assessing}. We believe that metamodels elements should be defined using domain concepts consistent with the application domain to (1) design understandable and consistent systems, (2) ease the communication with the many involved stakeholders, and (3) facilitate the maintenance of the systems. 

There have been a few research contributions that focused on the development of recommender systems for the auto-completion of models or metamodels. Early approaches have focused on logic and rule-based systems with model transformation to recommend partial models~\cite{Sen2007DSM_completion, Sen2007Partial_completion, Diagrammatic_completion}. More recent works have focused on the extraction of knowledge from the application domain through textual data and on the utilization of pre-trained word embedding models\footnote{
    In this paper, the term ``model'' refers to a machine learning model rather than an instance of a metamodel as customary in the MDE literature.
    All MDE artifacts we operate on are metamodels.
}~\cite{agt2018automated, burgueno2020nlp}. On the one hand, the early approaches require to manually design rules and define rather complex approaches. On the other hand, previous knowledge-based and natural language processing (NLP) approaches require the extraction of textual knowledge about the domain that are not always available in practice or that are rather tedious to extract with acuteness. 

In this paper, we propose a learning-based approach to recommend relevant domain concepts to a modeler during a metamodeling activity by training a deep learning model on thousands of independent metamodels. Our approach intends not only to be multi-objective by enabling the recommendation of concepts related to classes, attributes and references, but also to be easily adaptable to different end-tasks. Additionally, our approach neither requires to extract knowledge about the application domain nor hand-designed features or rules.
More specifically, we extract both structural and lexical information from more than $10\,000$ metamodels and train a state-of-the-art deep neural language model similar to BERT~\cite{devlin2019bert}. The trained model can then be used to recommend relevant domain concepts to the modeler given a specific modeling context. The goal of the system is to provide a relatively short-ranked list of suggestions that will help the modeler through the design of a metamodel.

We evaluate the proposed approach using 166 metamodels unseen during the training of our language model. The results of our experiments show that our trained model is able to learn meaningful domain concepts and their relationships among thousands of metamodels. It is also able to recommend relevant classes, attributes, and associations concepts in three modeling scenarios. In the two first scenario, we show that our model is able to recommend relevant domain concepts to rename elements of an already designed metamodel. In the second one, we show that our model can also be used to recommend domain concepts during the whole process of the design of a metamodel. Finally, we discuss our approach with both modeling scenarios using two specific metamodels and highlight interesting results and several opportunities to improve our approach. 

The contributions of this paper are as follows:
\begin{itemize}
    \item 
    A novel representation of Ecore metamodels encoding their structural and lexical information to learn domain concepts with deep learning models.
    
    \item 
    A publicly available repository containing encoded metamodels, all the data and our trained model.
    
    \item 
    The demonstration of the applicability of the approach to design intelligent modeling assistants for common modeling scenarios.
\end{itemize}

The rest of the paper is structured as follows. In \Sect{sec:pre_trained_lms}, we review the deep learning model that we use in this work. Next, we go through a motivating example and discuss how to transform the metamodels into trees to train our learning model. \Sect{sec:learning_domain_concepts} details our overall framework. \Sect{sec:experimental_setting} presents the experimental setting of our evaluation and the addressed research questions. We report the results of our experiments, the limitations of our approach and opportunities of improvements in \Sect{sec:results}. We discuss the related work in \Sect{sec:related_work}. Finally, we conclude and discuss future work opportunities in \Sect{sec:conclusion}.

\section{Pre-trained language models}
\label{sec:pre_trained_lms}
\begin{figure}[!t]
    \centering
    \includegraphics[width=\linewidth]{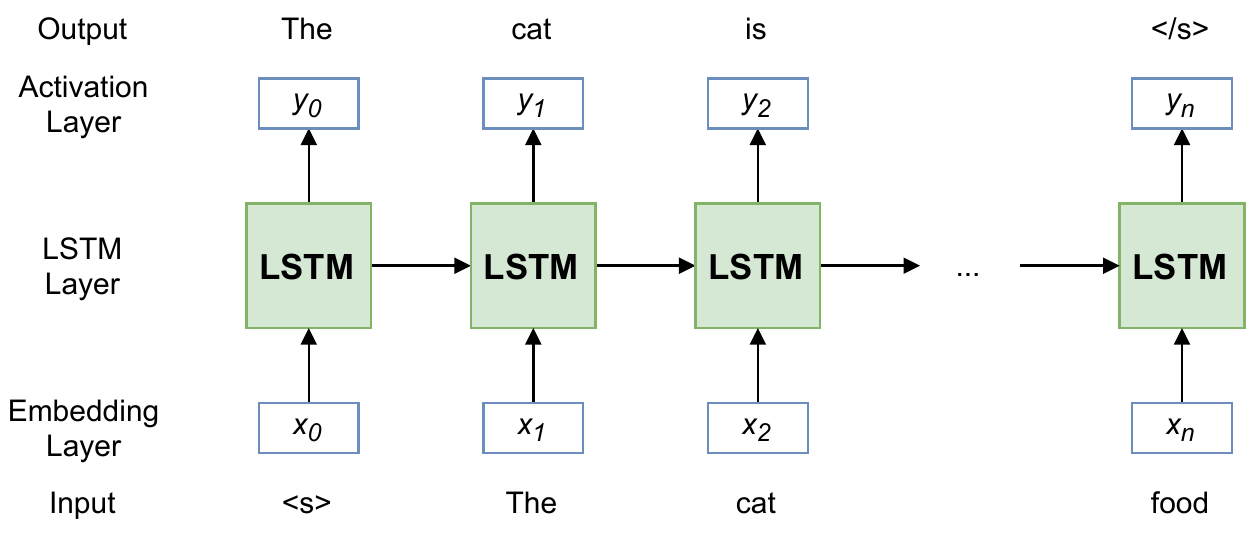}
    \caption{The architecture of a recurrent neural network with long-short term memory cells and with the input sentence: \texttt{<s>} The cat is eating some food \texttt{</s>}}
    \label{fig:lstm}
\end{figure}

Traditional language models such as \textit{recurrent neural networks}~\cite{nlm_2003} take textual data as input and try to predict the next word given the previous one as training objective (\ie causal language modeling). \Fig{fig:lstm} illustrates this architecture with \textit{long-short term memory cells} (LSTM)~\cite{lstm} which are widely used in the literature and helps the recurrent neural network to learn long-range dependencies from the input. Nowadays, the pre-trained language models architectures based on \textit{transformers and attention mechanism}~\cite{vaswani2017attention} are generally favored in a lot of situations as they result in state-of-the-art performance in a lot of tasks.

\begin{figure*}[!ht]
    \centering
    \includegraphics[width=.7\textwidth]{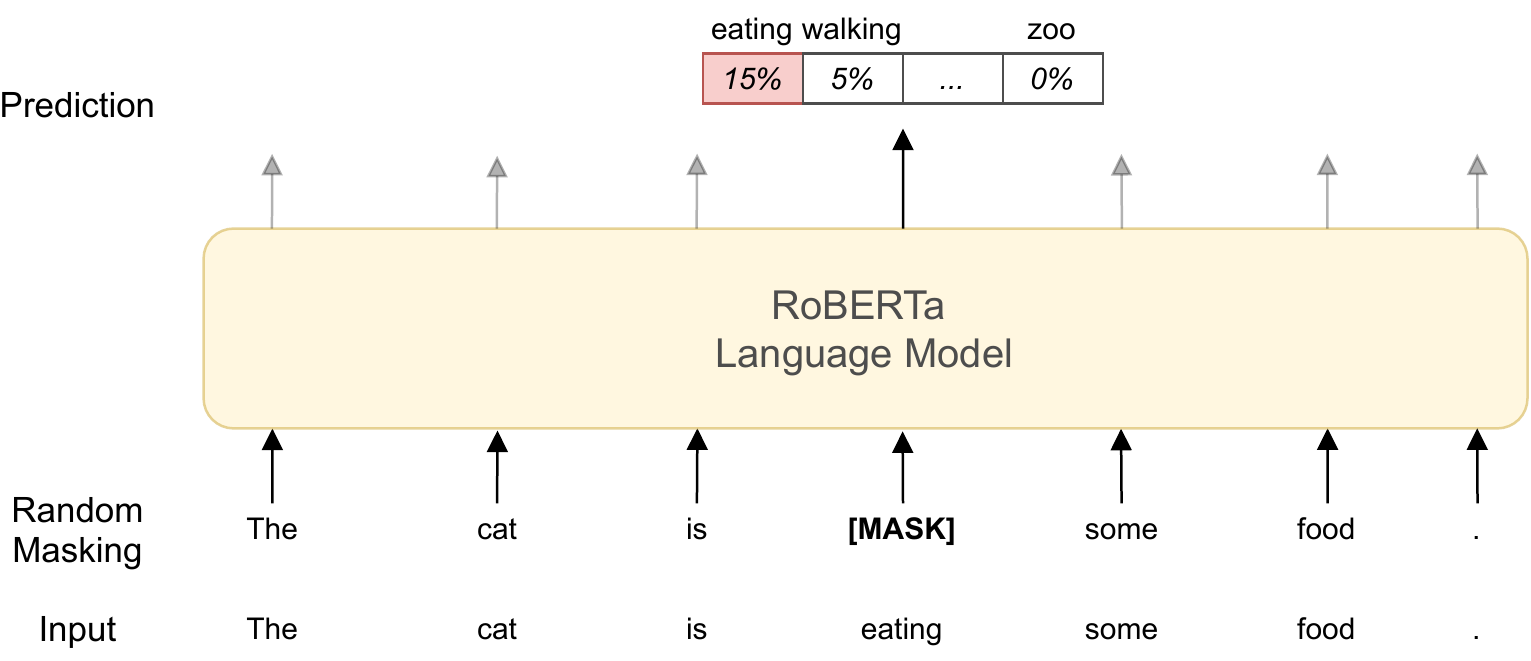}
    \caption{RoBERTa -- Masked language modeling with the input sentence: \textit{The cat is eating some food .}}
    \label{fig:roberta}
\end{figure*}

Over the last few years, progresses in natural language processing (NLP) have dramatically increased, in particular, thanks to the design of pre-trained bidirectional language models such as BERT~\cite{devlin2019bert}, GPT~\cite{Radford2018-GPT}, GPT2~\cite{radford2019-gpt2}, GPT3~\cite{brown2020language-gpt3}, XLM~\cite{xlm} or XLNet~\cite{XLNet}. They have the ability to learn high-level contextual representations of text in a self-supervised learning fashion. Besides, another advantage of these architectures is that one can adapt the neural network to a specific end-task by \textit{fine-tuning} it. 

In this work, we propose to reuse a pre-trained language model architecture without the pre-training. That is, we do not use an existing pre-trained language model and fine-tune it for our task. Instead, we train one using our data to learn contextual embeddings of metamodel concepts. More specifically, we use a \textit{RoBERTa} architecture and \textit{masked language modeling} as training objective~\cite{liu2019roberta}. 

RoBERTa is a language model architecture that improves BERT~\cite{devlin2019bert}. Its main objective is to reduce the computational costs of BERT and improve the masking algorithm. \Fig{fig:roberta} illustrates the masked language modeling objective\footnote{
  The illustration is inspired from the following blog: \url{http://jalammar.github.io/illustrated-bert/}
}.
The masked language modeling objective consists of masking a percentage of the input words (\textit{i.e.,} typically $15\%$). The objective for the model is to predict each masked word.
Additionally, RoBERTa architecture improves the pre-training data representation by tokenizing them using \textit{byte-level pair encoding} (\textit{similarly to GPT-2}). The advantage of such tokenization algorithm is that it produces an \textit{open-vocabulary} and allows encoding rare and unknown words. The usage of pre-trained language models in software engineering for program understanding tasks has shown to have some significant impacts thanks to their ability to learn powerful embedding spaces and to assist developers in complex tasks~\cite{kanade2019_pretrained,karampatsis2020scelmo_pretrained,feng2020codebert_pretrained}. 

\section{Motivation and metamodel representation}
\label{sec:motivation_metamodel_repr}

In this section, we first present a motivating example to illustrate the importance of developing recommender systems to assist modelers in the design of metamodels. Then, we discuss how to transform metamodels into structures that can be used to train language models presented in Section \textbf{\ref{sec:pre_trained_lms}}.

\subsection{Motivating example}
\label{sec:motivating_example}

\begin{figure*}
    \centering
    \includegraphics[width=0.7\textwidth]{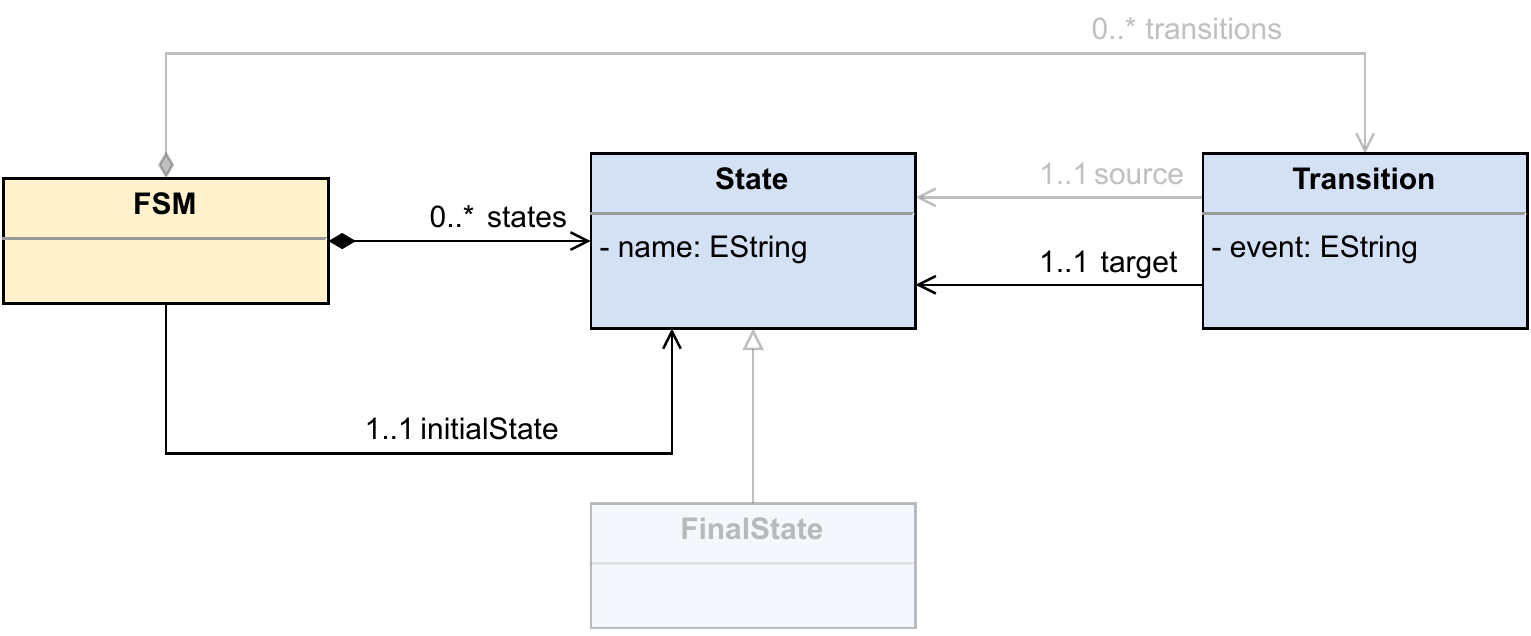}
    \caption{Motivating example --  finite state machine metamodel.}
    \label{fig:fsm}
\end{figure*}

To illustrate the rationale behind this work and the need for intelligent modeling assistants (IMAs), let us consider the motivating example in \Fig{fig:fsm} where a modeler is designing the metamodel of finite state machines (FSM). 

Let us also consider that the modeler is working in a modeling environment, such as Eclipse that provides a range of Ecore elements that can be dragged and dropped to build the metamodel. These elements are defined in the abstract syntax of Ecore and are manipulated through a concrete syntax as depicted in \Fig{fig:fsm}. Thus, such modeling environments are easily able to ensure the syntactical correctness of the metamodel under design by checking its conformance to the Ecore abstract syntax. However, the existing modeling environments generally do not incorporate tools to help the modeler to design pure semantic aspects of the metamodel. Here, we refer to the semantic as the ontological meaning of the metamodel but also as the meaning of each elements, the relationships among them and their meaning in the application domain.
Given the example in \Fig{fig:fsm}, the ontological meaning of the metamodel can be mapped to a concrete description in the application domain of FSMs. Each single element of the metamodel has a meaning that depends on its instantiation in the metamodel. For example, a FSM is composed of states which can eventually be final states. The concept of ``State'' depends on how it is defined and how it interacts with other elements of the metamodel. In practice, a modeler defines implicitly the semantic of the metamodel by instantiating domain concepts (\eg the concept of FSM or State) and defining how these concepts should be related. 

Manipulating these domain concepts in an adequate way is tedious for the modeler as (1) she may not be an expert of the application domain, (2) metamodels can rather become complex, and (3) the concepts she manipulates during a modeling activity must be handled so that the metamodel embodies correct semantics for the application domain. In the FSM example, it makes sense to have a class called ``\textit{Transition}'' that references another class ``\textit{State}'' via two associations:  one for the source and one for the target state of the transition. In fact, all the aforementioned concepts are well-defined in the theory of FSM and the names chosen in the provided example are consistent since they refer to concepts coming from the application domain. Additionally, for more complex metamodels that could either become large or mix application domains, it is crucial to have coherent and conceptually consistent usage of domain concepts to guarantee a certain quality of the metamodel. 

Therefore, our goal is to alleviate the modeler's burden by developing an IMA that could provide relevant recommendations of domain concepts for a given metamodel. In fact, we envision that both structural and textual regularities can be abstracted from existing metamodels using a NLP learning model as follows: (1) the structural regularities can help the model to determine what type of metamodel element is likely to come after a specified context and (2) the model textual regularities can help the model to determine a meaningful embedding space which consists of a vector space that  encodes semantic properties of the metamodel domain concepts. These regularities can then be leveraged to build on a recommender system that is able to adapt itself to the current modeling context, \ie by providing relevant recommendations, ranked by likelihood, given a metamodel under design. 


\subsection{Metamodels as trees}
\label{sec:ecore-to-tree}

Since we rely on a NLP learning model, the data needs to be available in a textual format.
We propose a way to extract meaningful information from the metamodels by transforming them into trees that capture both lexical and structural information about the metamodels. In view of the target modeling tasks, we extract metamodel elements that support the task(s) while ignoring elements that may not be relevant and that could introduce noise during the training of our model.

\begin{figure}[!t]
    \centering
    \includegraphics[width=\linewidth]{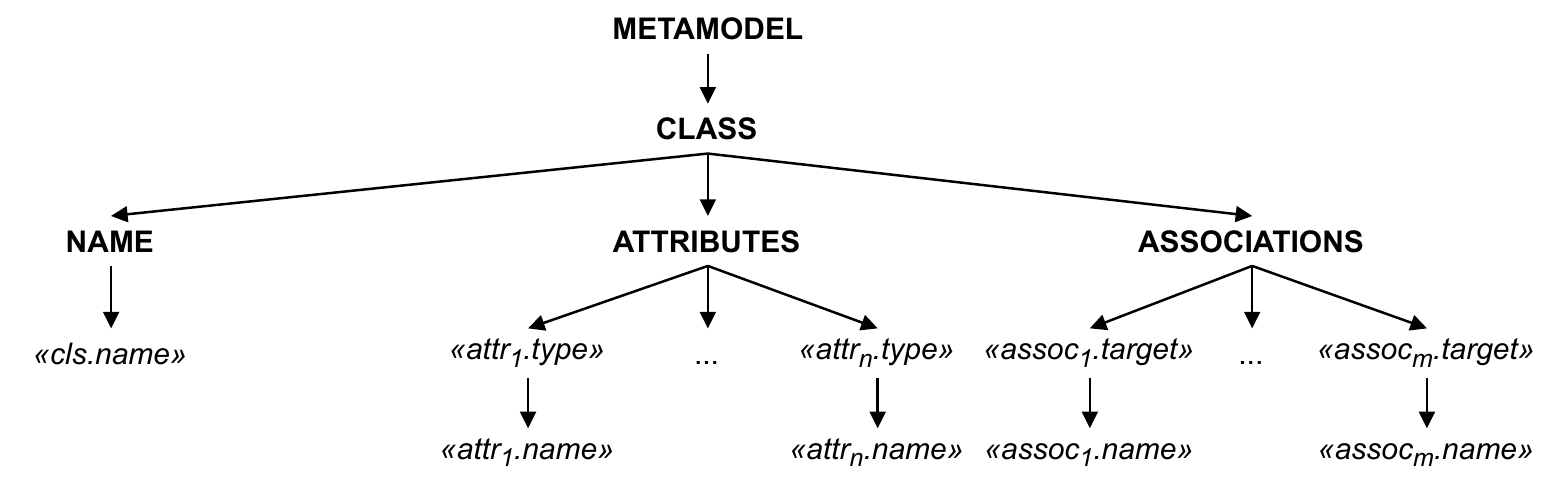}
    \caption{General structure of a metamodel tree}
    \label{fig:tree_general}
\end{figure}
\begin{figure}[!t]
    \centering
    \includegraphics[width=\linewidth]{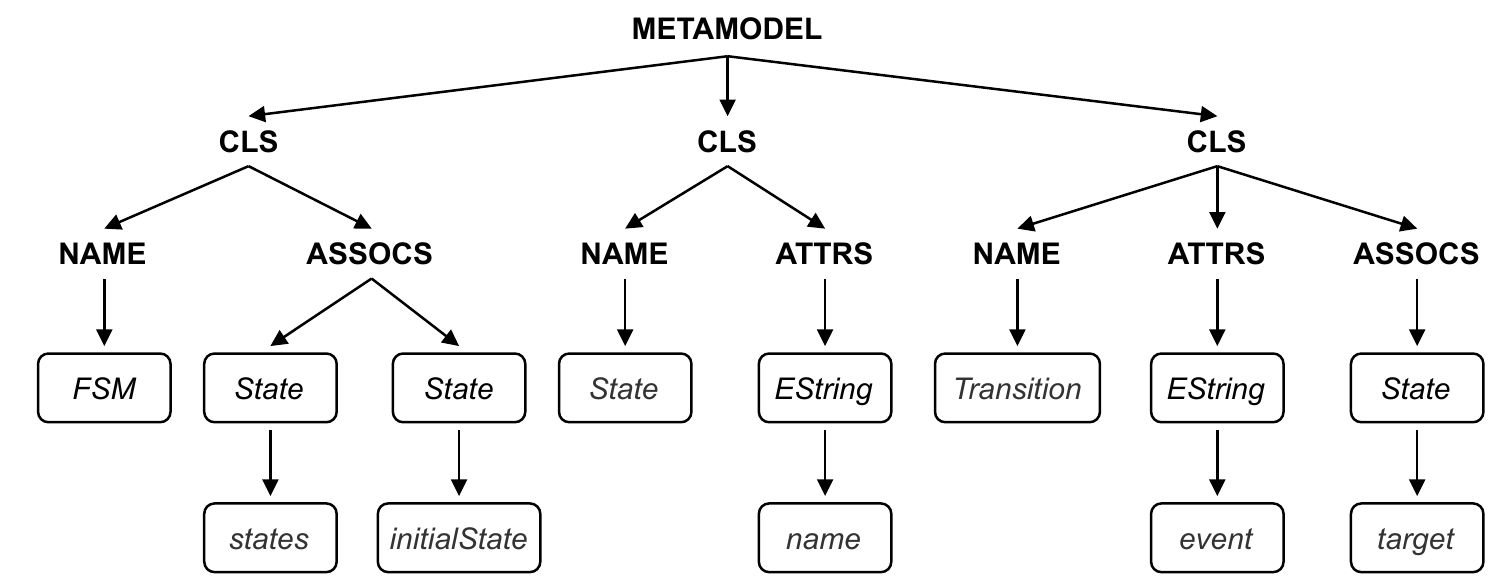}
    \caption{Tree structure of the partial FSM metamodel.}
    \label{fig:tree_example}
\end{figure}

\Fig{fig:tree_general} depicts the general tree structure of a transformed metamodel.
The transformation assumes that a metamodel contains classes, their attributes and the associations.
The resulting tree is rooted by a generic node representing the metamodel.
Then, each class in the metamodel defines a sub-tree containing its own information rooted in a \Code{CLS} node. From left to right, we first have a node \Code{NAME} that references a leaf node with the class name. Secondly, the \Code{ATTRS} node contains two nodes for each attribute of the class: one for its type and the leaf is for its name. Similarly, the \Code{ASSOCS} node contains nodes representing all associations (including compositions) with the current class as source. We create two nodes per association: one for the name of the target class and the other for the name of the association. For both attributes and associations, we also encode their types to provide additional semantic information to the learning model. The rationale for defining this tree structure is that it specifies a hierarchy between metamodel elements that can be learned by a language model. \Fig{fig:tree_example} shows the tree structure of the non-shaded part of the metamodel in \Fig{fig:fsm}. This tree representation is similar to the one used by Burgueno \etal{}~\cite{burgueno_lstm} with which the authors learned model transformations using neural networks. Our tree-based representation differs from theirs as they transform instances of models into tree representations that encode information such as the values of the attributes that are instance-specific. Note that our representation does not capture composition hierarchies explicitly.

Our approach does not depend on how trees are defined. For instance, it is possible to ignore the target of associations or to include generalization relations. Additionally, the order in which elements appear in the tree could be altered. In this work, we choose to work with the ``natural'' order of elements as they appear in the metamodel. If, for example, metamodels are provided in Ecore, we assume that the order in which their elements appear in Ecore file reflects (partially) the order in which a modeler would have designed and saved the metamodel. In addition to that, even though we present a tree-based representation for MOF-like metamodels, our approach can generalize to other kind of metamodels which can be expressed as relational graphs.

Finally, the tree representation of the metamodel allows us to train a language model by flattening its structure into a textual sequential representation. We discuss this aspect of the training phase in the next section.

\section{Learning metamodel domain concepts}
\label{sec:learning_domain_concepts}
To tackle the problem of recommending relevant domain concepts to the modeler during a metamodeling activity, we envision that the high-level concepts manipulated during such an activity (\textit{e.g.,} the concept of FSM in \Fig{fig:fsm}) can be abstracted from a dataset of designed metamodels.
Deep learning, and particularly self-supervised and unsupervised learning techniques, learn representations that encode semantic information about the world: in our case, metamodels.
We foresee that pre-trained language models can be leveraged to learn domain concepts of metamodels. Such a trained learning model can be used to perform predictive modeling task to help a modeler to design conceptually sound metamodels. In this work, we present our approach on two practical applications, but our model can easily be adapted for other tasks by fine-tuning it.

\begin{figure*}[!t]
    \centering
    \includegraphics[width=\textwidth]{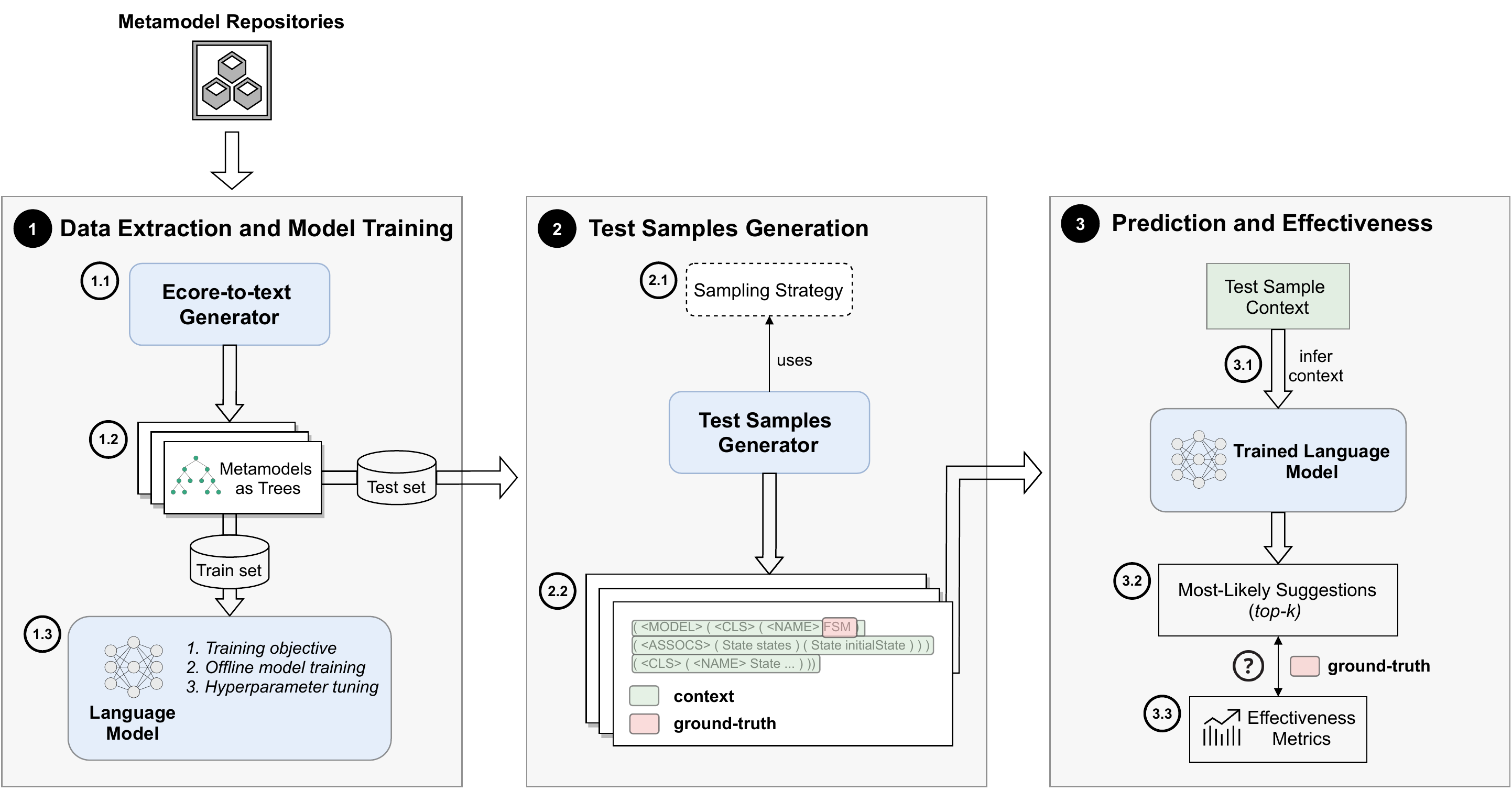}
    \caption{Overall framework of the approach.}
    \label{fig:approach}
\end{figure*}

In this section, we present a framework to learn the aforementioned domain concepts and exploit them to recommend relevant concepts to the modeler by leveraging contextual information specific to the current modeling activity. \Fig{fig:approach} depicts the 3-phased architecture of our approach. We explain each phase in what follows.

\subsection{Data Extraction and Model Training}
\label{sec:data_model_training}

In order to train efficiently pre-trained language model, we need to gather data relevant to the target objective. In general, the more data we have, both in term of diversity and quantity, the better it is to allow the model to generalize on unseen examples. Model repositories~\cite{mdeforge, atlanmod, remodd} are rich with metamodels, often stored in XMI, that can be crawled. In \Fig{fig:approach}, the crawled metamodels enter the pipeline through the \textit{Data Extraction and Model Training} process.

\begin{figure*}[!ht]
    \centering
    \includegraphics[width=.7\textwidth]{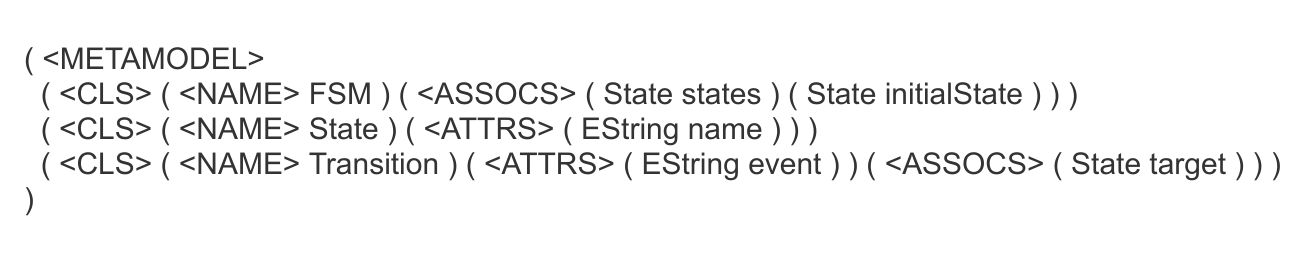}
    \caption{Textual tree structure of the partial FSM metamodel (see \Fig{fig:fsm}).}
    \label{fig:metamodel_tree_text}
\end{figure*}

Given the metamodels in a serialized format, the first step consists of transforming these metamodels into a format that can be given as input to a language model. In order to achieve this step, we define in \circled{1.1} an \textit{Ecore-to-text Generator} component. The role of this component is to transform metamodels into textual tree structures as presented in Section \textbf{\ref{sec:ecore-to-tree}}. The most straightforward way to perform this transformation is to define a model-to-text transformation. For each metamodel, the transformation outputs its corresponding tree structure representation \circled{1.2} as depicted in \Fig{fig:metamodel_tree_text} for the non-shaded part of the FSM metamodel presented in \Fig{fig:fsm}. Even though the language model takes sequential data as input, such a representation allows the model to learn from both structural and lexical information contained within a metamodel. That is, the structural aspect of the metamodel can still be implicitly learned by the language model as the textual format incorporates syntactical elements that allow it to contain hierarchical and structural information of the metamodel. Also, our approach could be used with the text-based representation of metamodels proposed as part of the Eclipse Emfatic framework~\cite{emfatic}.

Next, the metamodels are separated into a training set and a test set. The latter is left aside for the next steps of the approach. Before training a model, the data can pass through a NLP pipeline that could include several processes such as text normalization or tokenization. In practice, applying tokenization techniques such as subword tokenization or byte-pair encoding~\cite{sennrich-bpe} before training a language model helps reducing the size of the vocabulary and produces an open-vocabulary that provides the ability to a trained model to generalize better at testing time. At step \circled{1.3}, we train a language model using the training set in an \textit{offline} mode. Prior to training, we define a \textit{training objective} that the language model will try to fulfill during the training in order to update its parameters and learn from the data. For instance, the training objective could be a masked language modeling objective similar to BERT~\cite{devlin2018bert} or a permutation language modeling objective used in XLNet~\cite{yang2019xlnet}. Finally, the hyperparameters of the language model are tuned using either a validation set extracted from the training set or $k$-fold cross-validation.

\subsection{Test Samples Generation}
\label{sec:test_generation}

Before evaluating the trained model on a useful modeling task, we define a process which aims at generating test samples for the defined target task. Here, we emphasize the fact that the modeling task should reflect an actual use case that makes sense in the real world in order to evaluate the effectiveness of the trained model in a relevant modeling scenario. This important aspect is materialized by the \textit{Sampling Strategy} \circled{2.1} which must be conceived in a thoughtful way to concretise a modeling scenario as close as possible to a real-world use case. In fact, as we will see further, the sampling of a particular test data point corresponds to a state of a metamodel at a specific time. Depending on the modeling task, the sampling is not always straightforward and requires some thought.

\begin{figure*}
    \centering
    \includegraphics[width=.7\textwidth]{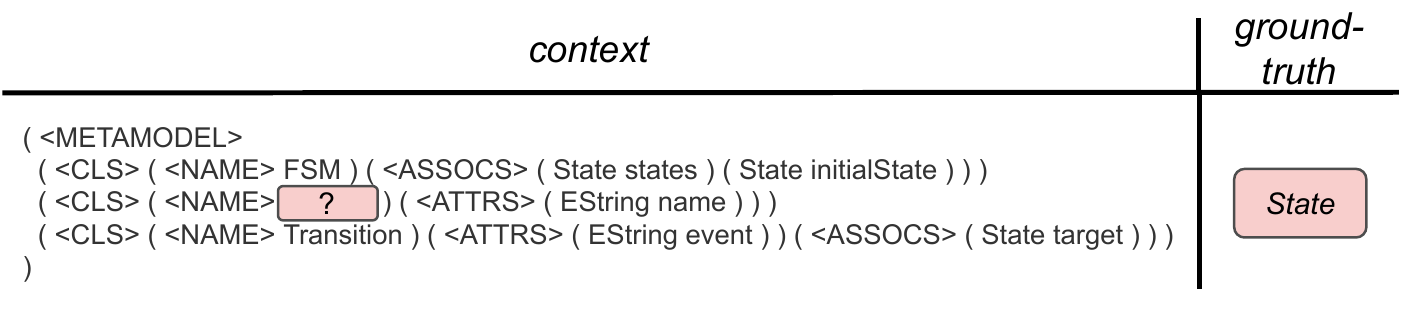}
    \caption{Excerpt of the generation of a test sample in a renaming scenario.}
    \label{fig:sampling_strategy_example}
\end{figure*}

Given a sampling strategy, the \textit{Test Samples Generator} generates test samples using the test set that corresponds to the strategy \circled{2.2}. For the sake of clarity, we showcase one possible sampling strategy in \Fig{fig:sampling_strategy_example}. This particular sampling represents a \textit{renaming modeling scenario}. That is, given a fully-designed metamodel, we obfuscate one element of the metamodel (\ie ground-truth) and consider all the other elements as its context. For a trained model, the goal would be to predict the ground-truth considering that context as input. In the real-world, this scenario could be considered to rename elements of a metamodel that do not represent well the manipulated concepts for a particular application domain. In \Sect{sec:experimental_setting}, we describe three modeling scenarios on which we evaluate our approach.

\subsection{Prediction and Effectiveness}
\label{sec:prediction_effectiveness}

The last process of our framework concerns the evaluation of a trained language model. For the model to be able to provide meaningful recommendations, we need to leverage the contextual information of each test sample, as discussed previously. Given a test sample, we provide its context to the language model \circled{3.1}. Then, the model outputs a list of candidates that are likely to match the obfuscated element (\ie ground-truth). At step \circled{3.2}, this list is retrieved using a parameter $k$ that determines the size of the list. Finally, at step \circled{3.3}, we compute \textit{effectiveness metrics} to evaluate the ability of the model to provide relevant recommendations given the current modeling scenario. 

To summarize, the trained model can be seen as a recommender system that takes as input contextual data about the modeling activity and outputs a ranked-list of most-likely recommendations that are assessed through the usage of widely-used effectiveness metrics.

\section{Experimental setting}
\label{sec:experimental_setting}
To evaluate our approach, we articulate our research questions around three simulated real-world modeling scenarios.
The goal of our experiments is the following:

\begin{oframed}
\noindent\textbf{Goal:}
    Determine to what extent a state-of-the-art pre-trained language model can provide meaningful domain concepts recommendations to a modeler during a modeling activity to reduce its cognitive load and ease the design of complex metamodels. 
\end{oframed}

In this section, we formulate the research questions and follow with detailed discussions about our dataset, how we addressed the research questions, the model training and the evaluation metrics.

\subsection{Research Questions}\label{sec:rqs}

The experiments are divided into two parts.
Each of the following research question aims at assessing the usefulness of the aforementioned model in a distinct modeling scenario:
\begin{itemize}
    \item \textbf{RQ1 -- Scenario 1:} \textit{Given an already designed metamodel, is our approach able to recommend relevant domain concepts to rename metamodel elements?} \\
    In this scenario, we assume the metamodel if completely defined. The objective is then to evaluate our model on its capability to recommend useful domain concepts to rename specific elements of the metamodel such as classes, attributes, and associations by considering all of the information of the metamodel as a global context.
    \\
    
    \item \textbf{RQ2 -- Scenario 2:} 
    \textit{Does considering a local context made of the metamodel elements close to the one being renamed increases the effectiveness of the system?} \\
    This scenario is similar to the one in RQ1 except that the sampling strategy differs. By considering a global context we may introduce noisy contextual data that could reduce the precision of our model. Therefore, we investigate whether considering a local context made only of metamodel elements close to the one being renamed can improve the effectiveness of our model.  \\
    
    \item \textbf{RQ3 -- Scenario 3:} 
    \textit{While incrementally designing a metamodel, is our approach able to recommend relevant domain concepts at each increment, while maintaining a prediction context made of the elements in a previous increment?} \\
    This modeling scenario is substantially more complex than the previous one. It aims at simulating how a modeler designs a complete metamodel in a modeling editor. Here, sampling of the test data is more complex. It allows us to report on the evolution of the effectiveness of our model as the metamodel is designed progressively.
\end{itemize}

The last research question provides an in-depth evaluation of our approach on two specific metamodels.
\begin{itemize}
    \item \textbf{RQ4 -- Use Cases:} 
    \textit{What are the advantages and the limitations of the latter?} \\
    We choose one domain-specific and one general purpose metamodel in our test set and perform a qualitative analysis to identify specific cases where our approach performs well or not.
    We highlight its limitations as well as some opportunities of improvement. 
\end{itemize}

\subsection{Datasets}
\label{sec:dataset}

We used the MAR dataset\footnote{
  \url{http://mar-search.org/experiments/models20/}
}
which contains around $17\,000$ Ecore metamodels crawled from Github and AtlanMod Zoo~\cite{mar_search_engine}. The former consitute the training dataset and the latter is the test set. Before extracting tree-structured data from the metamodels, we remove metamodels containing less than 2 and more than 15 classes. With too few classes, our pre-trained language model would not be able to learn meaningful information from the metamodels due to a lack of context. Conversely, two many classes would lead to considering very large trees, resulting in an intractable and computationally costly training phase. We end up with $10\,000$ metamodels for training, $1\,112$ for the validation of the hyperparameters of the pre-trained language model and $166$ for testing.

\begin{table}[!t]
\caption{Training set used in the experiments. Identifiers corresponds to the total number of classifiers, attributes and associations in the dataset. Types is the number of unique identifiers. Hapax Legomena is the number of identifiers that appear only once in the dataset.}
    \centering
    \small
    \begin{tabular}{cccc} \toprule
        & \# Identifiers & Types & \# Hapax Legomena \\ \midrule
        \textbf{Train Set} & 267.370 & 58.027 & 31.911 (55\%) \\ \bottomrule
    \end{tabular}
    
    \label{tab:training_stat}
\end{table}

After filtering the dataset, we extract trees as described in Section \textbf{\ref{sec:ecore-to-tree}}. We use Xtend\footnote{https://www.eclipse.org/xtend/} combined with Eclipse EMF API\footnote{
  \url{https://www.eclipse.org/modeling/emf/}
}
to programmatically transform the metamodels into trees using a model-to-text transformation. \Table{tab:training_stat} reports the number of identifiers (i.e., classes, attributes, and associations), the number of types (\ie unique identifiers) and the number of hapax legomena (\ie identifiers that appear only once). As we can observe, we have a high proportion of hapax legomena (\textit{i.e.,} more than $50\%$ of the unique identifiers) in the resulting dataset.
Considering those identifiers that appear very rarely in a corpus could lead to learning a lot of noise.
To cope with this issue, we apply a \textit{byte-pair encoding} tokenization algorithm~\cite{sennrich-bpe}. The objective of this process is to decompose identifiers into \textit{subword units} (or \textit{merges}). The tokenizer keeps the most-common merges in a vocabulary. Thanks to byte-pair encoding, the pre-trained language model is able to generate identifiers unseen during training or rare identifiers by composing from the merges.

\begin{figure}[!t]
    \centering
    \includegraphics[width=1\linewidth]{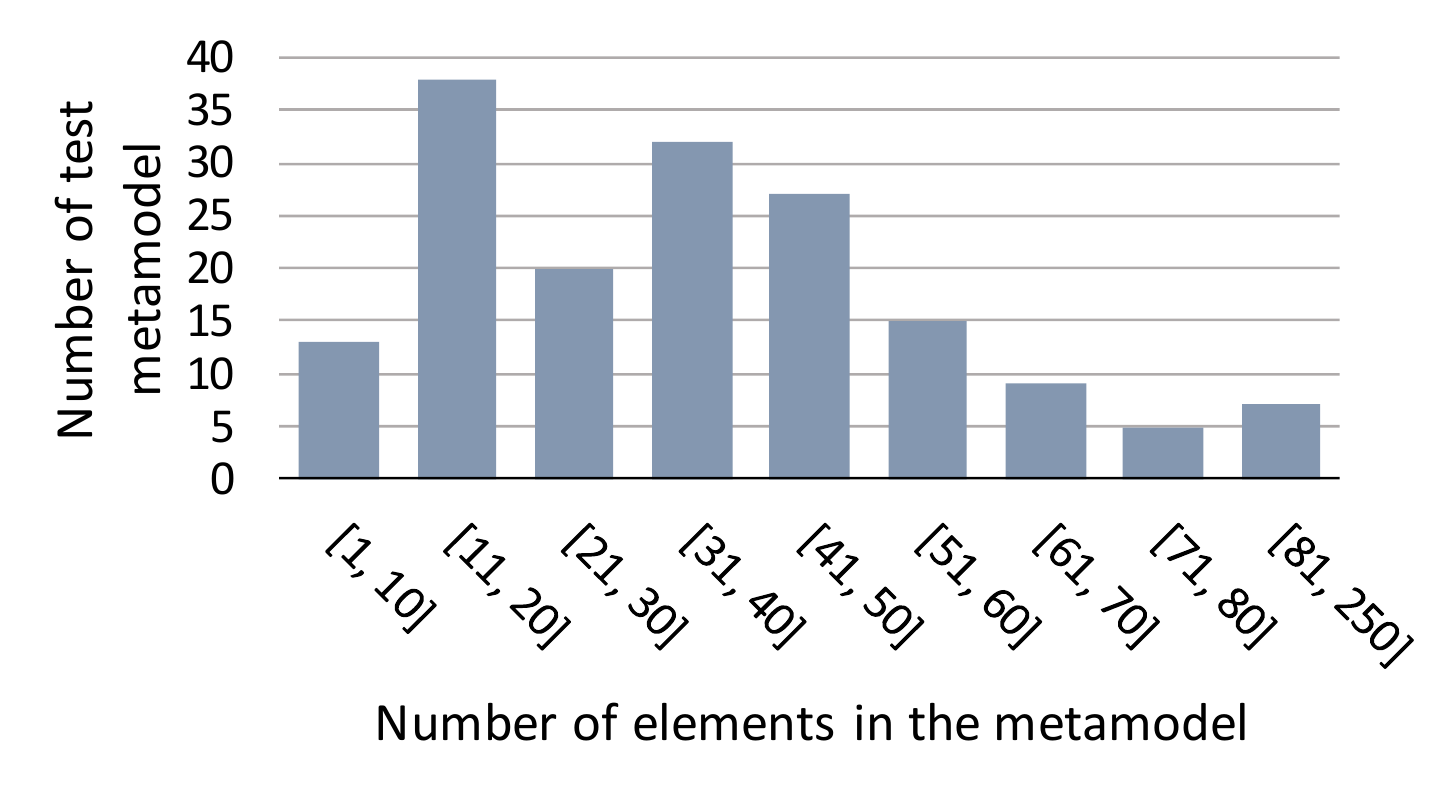}
    \caption{Distribution of the metamodels in the test set per intervals of number of elements}
    \label{fig:test_set_stats}
\end{figure}

In \Fig{fig:test_set_stats}, we report the distribution of our test metamodels in term of their number of elements, where an element is either a class, an attribute, or an association. This figure shows that our test set is diverse, which enables the evaluation of our approach in a various range of situations. 

\subsection{How we address RQ1 -- RQ2}
\label{sec:how_rq1_rq2}

Both RQ1 and RQ2 rely on a common end-task which is about renaming metamodel elements. For instance, the task could make sense in a context where a metamodel would have been defined with rather vague concepts. In this particular situation, we believe that our approach could be valuable to help the modeler to rename the elements of the metamodel in a meaningful way. 

We refer to the sampling strategy of RQ1 as a \textit{global context sampling} because we sample each test data (i.e., an element to be renamed) by considering the whole metamodel as its context. We discussed this sampling in \Sect{sec:test_generation} as one possible sampling strategy that can be applied in our approach.

\begin{figure}[!t]
    \centering
    \includegraphics[width=.9\linewidth]{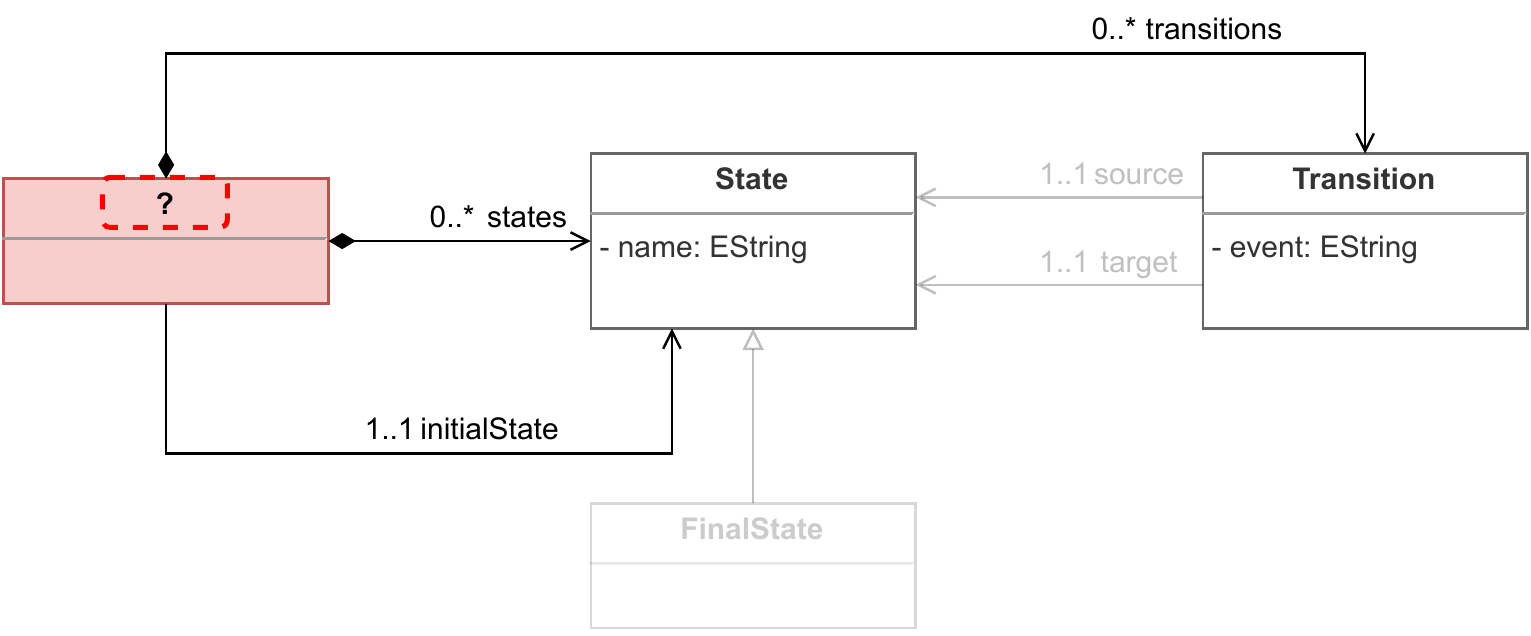}
    \caption{Example of the \textit{local context sampling} strategy.}
    \label{fig:ex_local_sampling}
\end{figure}

Alternatively, we refer to the sampling strategy of RQ2 as a \textit{local context sampling}. The rationale for investigating this research question is that we conjecture that a global context sampling does not allow the approach to scale to medium-large metamodels. Considering a global context could include elements of the metamodels that are somehow unrelated to the element of interest and could thus result in imprecise predictions. As a consequence, we investigate whether a local context sampling could increase the effectiveness of our pre-trained language model by limiting the context of the element to be renamed to only elements that are close to it in the metamodel. In \Fig{fig:ex_local_sampling}, we illustrate this sampling strategy on the FSM example. The context corresponds to the non-shaded part of the metamodel and the mask is ``FSM''. The local context contains the metamodel elements that are directly connected with an association to the element that needs to be predicted. In the case of attributes, we keep the elements that are linked with the class of the attribute in the context.

\subsection{How we address RQ3}
\label{sec:how_rq3}

\begin{figure*}[!t]
    \centering
    \begin{subfigure}[t]{.3\textwidth}
        \centering
        \includegraphics[width=.4\linewidth]{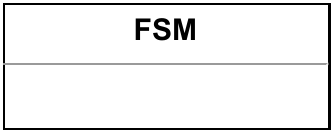}
        \caption{}
    \end{subfigure}
    \vline
    \hfill
    \begin{subfigure}[t]{0.3\textwidth}
        \centering
        \includegraphics[width=\linewidth]{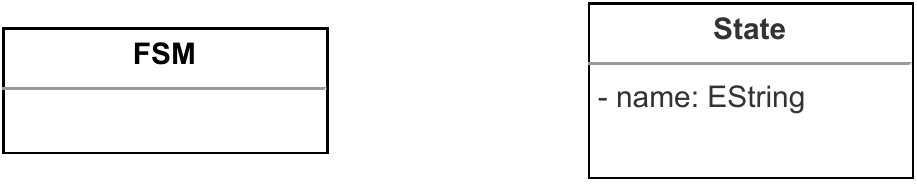}
        \caption{}
    \end{subfigure}
    \hfill
    \vline
    \hfill
    \begin{subfigure}[t]{0.3\textwidth}
        \centering
        \includegraphics[width=\linewidth]{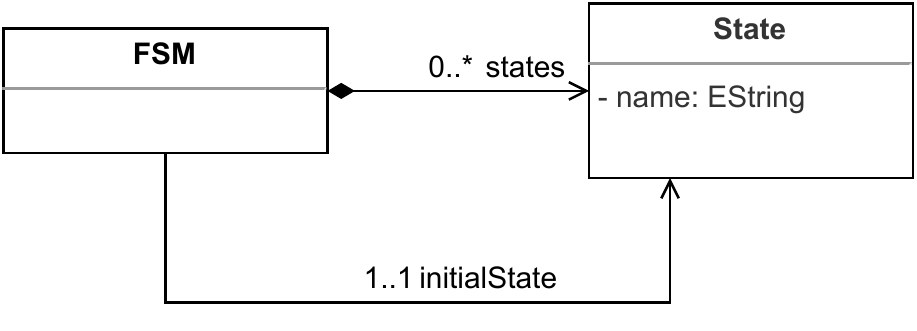}
        \caption{}
    \end{subfigure}
    
        \vspace{\floatsep}
    \begin{subfigure}[t]{0.4\textwidth}
        \centering
        \includegraphics[width=\linewidth]{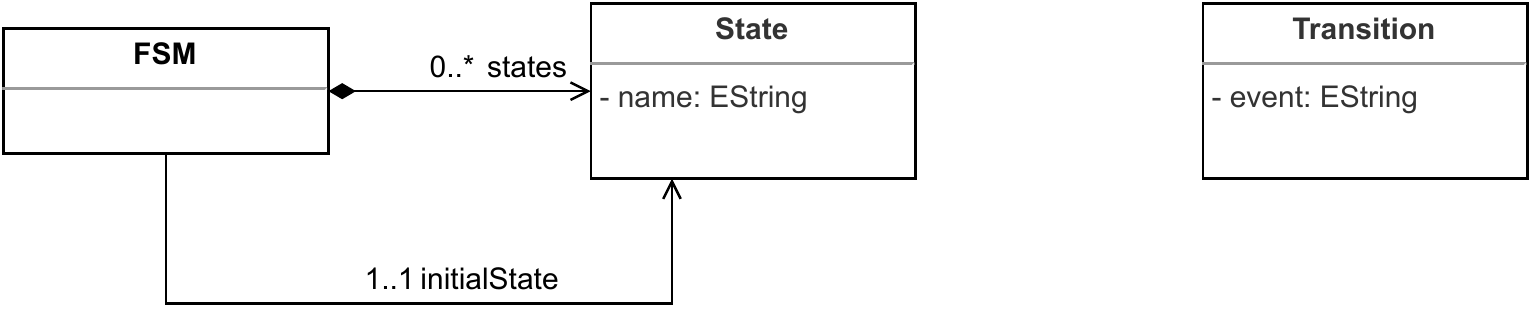}
        \caption{}
    \end{subfigure}
    \hspace*{2em}
    \vline
    \hspace*{2em}
    \begin{subfigure}[t]{0.4\textwidth}
        \centering
        \includegraphics[width=\linewidth]{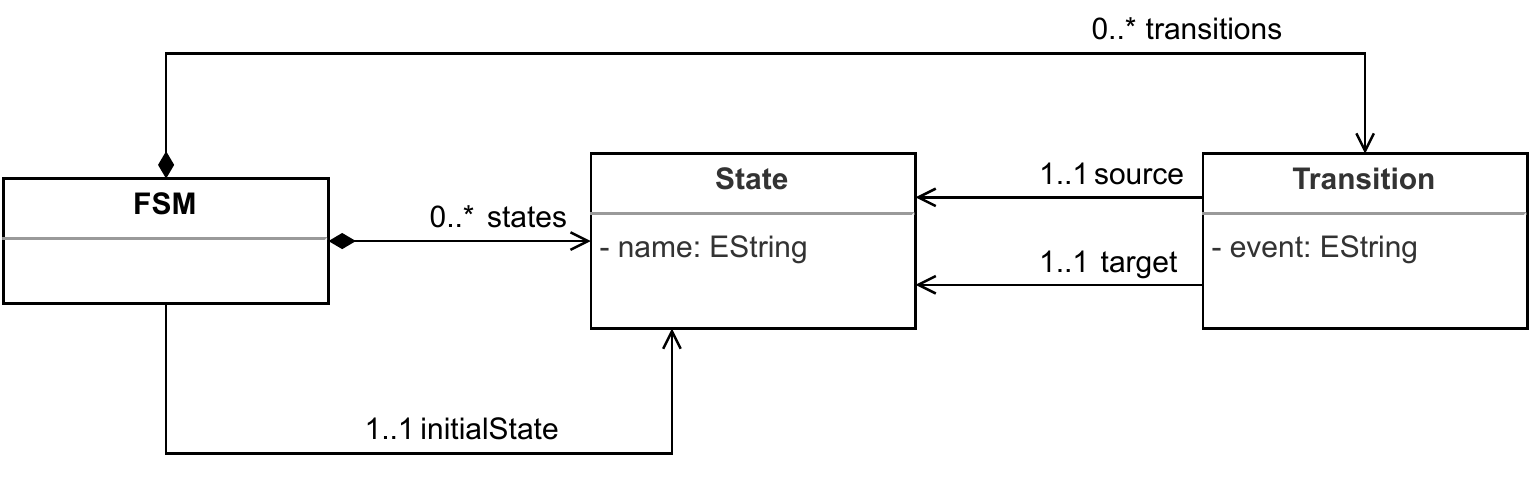}
        \caption{}
    \end{subfigure}
    \caption{Illustration of the constructive test sampling}
    \label{fig:rq3_sampling_example}
\end{figure*}

To address RQ3 thoroughly, we seek to simulate the construction of the metamodel in a real modeling environment. For instance, we take into account the fact that two classes must exist before a modeler creates an association between them. \Fig{fig:rq3_sampling_example} illustrates the sampling on the FSM metamodel.

To initiate the sampling, we start with an empty metamodel and first retrieve a \textit{root class}, \ie one that does not have any incoming association (like the FSM class). If the metamodel does not contain such an element, we choose the class that has the least number of incoming associations as root.
Note that the root class is not predicted as our model requires contextual data to provide recommendations.
We set the root class as our current class. The rest of the sampling process is iterative:
\begin{enumerate}
  \item Iterate over all the class that are linked to the current class and choose one randomly.
  \item Generate a test sample where the ground-truth is the name of the chosen class and generate one test sample for each of its attributes.
  \item Create one test sample for each association that links the current class and the chosen one.
  \item Choose the next current class that can be reached (by an association) from the current one. If none can be reached, the next current class is unvisited yet and has the least number of incoming associations.
  \item Loop back to step 1 and stop the process when the metamodel is fully re-constructed.
\end{enumerate}

As we progress in the construction of the metamodel, the context that can be used to predict the next element increases. In this scenario, we consider the context to be all the elements that were previously constructed. Therefore, we suspect that the effectiveness of the trained model will increase as the construction evolves thanks to the growth of the context that provides more contextual information about the current modeling activity. Nonetheless, we emit certain reserves on this assumption as the growing size of the context may also introduce noise as it may contain metamodel elements far from the one being that do not contribute to make the model accurate given a current modeling context.

\subsection{How we address RQ4}
\label{sec:how_rq4}

The objective of RQ4 is to report qualitative results on two conceptually different metamodels drawn from our test set. The first one is a \textit{Petri nets} metamodel and the second is a \textit{Java} metamodel. The former uses rather broad concepts that seem to be less domain-specific and more widely-used in the definition of common metamodels than the latter metamodel. Therefore, this RQ allows us to compare the effectiveness of our approach depending to which extent the test metamodel is domain-specific. 

We report several top-$5$ recommendations lists provided by our trained model for both metamodels to showcase the ability of our model to recommend concepts close to the application domain of the test metamodel. We chose to report top-$5$ recommendations as it allows us to analyze qualitatively the model in a quiet restrictive scenario and as reporting more recommendations does not affect the conclusions of the evaluation. To achieve this, we use the modeling scenario of RQ1 -- RQ2 that gives us the best performance. Finally, we evaluate the scenario of RQ3 on both metamodels.

\subsection{Model Training}
\label{sec:model_training}

Pre-trained learning models contain various hyperparameters that need to be optimized to avoid phenomena such as overfitting the data or low generalization performances. To train our models, we divided our training set into a training and validation set with a 90/10 ratio. We monitored the model losses on both training and validation sets after each training iteration, where one iteration means that the model has gone through all the training samples. The losses enable us to monitor the evolution of the training of the model over the iterations and help us determining when to stop the training phase to avoid phenomena such as overfitting. Due to the wide range of hyperparameters and the complexity of the model, we were not able to optimize all of them. We report the hyperparameters values in Appendix \ref{sec:model_hp}.

We trained a RoBERTa language model~\cite{liu2019roberta} using a masked language modeling objective with Huggingface's implementation~\cite{wolf2020huggingfaces}. We use a Linux-Gentoo Server with 4 x Nvidia GeForce RTX 2080 GPUs, an AMD Ryzen Threadripper 3970X CPU and 126 GB of RAM. The training is independent from the testing phase as it is done offline and the optimal model is saved on a hard disk. 

During the training phase, the model masks randomly 15\% of the tokens in the input and attempts to reconstruct them (\ie masked language modeling objective). We did not \textit{fine-tune} the model for any scenario, \ie during the training phase, the model does not learn how to adapt itself to each specific scenario. This fine-tuning step can be achieved by resampling all the training sample according to each specific scenario and resuming the training from a previous checkpoint. Additionally, the model we provide can be fine-tuned for other tasks that rely on different objectives. For instance, it could be fine-tuned on a \textit{translation objective} that translates metamodel requirements into an actual metamodel. We discuss some of these opportunities in the future work section.


\subsection{Effectiveness Metrics}
\label{sec:metrics}

We measure the performance of our model on predictive tasks that aim to provide a relevant recommendation list for a given target. The recommendation is a list of strings ranked in decreasing order of relevance. To evaluate quantitatively our model, we use two commonly used metrics in very similar tasks such as code completion~\cite{Karampatsis_2020,svyatkovskiy2020fast,weyssow2020combining}, namely the Recall@$k$ and the Mean Reciprocal Rank (MRR). We consider a recommendation to be correct only if it matches exactly the ground-truth.

The Recall@$k$ for a set of test samples is the number of times the ground-truth (\ie target) appears in a recommendation list of size $k$, on average. Therefore, a greater value of Recall@$k$ is preferable as the system would make more accurate recommendations. 

The MRR for a set of test samples of size $T$ is the inverse of the rank of the correct recommendation in the list, on average:
$$
    MRR = \frac{1}{|T|} \sum_{i=1}^{|T|} \frac{1}{rank_i}
$$

The MRR is complementary to the Recall@k as it gives an idea of the ability of the model to suggest relevant tokens at the top of the list. For example, if on average, the correct recommendation appears at rank 2, the MRR is 0.5. Thus, a MRR close to 1 results in a perfect system.

Note that in all our experiments, the modeling tasks are rather complex and restrictive for the model as it has to predict the complete ground-truth and we do not consider a partially correct recommendation to be partially correct. As a consequence, only recommendations that contain the complete ground-truth have a positive impact on the effectiveness of the system.

\section{Results}
\label{sec:results}
\begin{table*}
\caption{[RQ1 - RQ2] -- Results in term of Top-1 score, Recall@k and MRR@k.}
\begin{center}
    \renewcommand{\arraystretch}{1.4}
    \setlength{\arrayrulewidth}{.5pt}
    
    \begin{tabular*}{\textwidth}{@{\extracolsep{\fill}}*{10}{ll|ccccccc|c}}
    \toprule
        \multicolumn{2}{l}{} &
        \multicolumn{8}{c}{\textbf{Metrics}} \\
        \cline{3-10}
    
        \multicolumn{1}{l}{\textbf{Scenario}} & \multicolumn{1}{c}{\textbf{Test}} & \multicolumn{1}{c}{Top-$1$} & \multicolumn{1}{c}{R@5} & \multicolumn{1}{c}{MRR@5} & \multicolumn{1}{c}{R@10} & \multicolumn{1}{c}{MRR@10} & \multicolumn{1}{c}{R@20} & \multicolumn{1}{c}{MRR@20} & \multicolumn{1}{c}{Size}  \\ 
        
    \midrule
        \multirow{3}{*}{\textbf{Full Context}} & Classes & 39.79 \% & 45.45 \% & 0.42 & 47.97 \% & 0.42 & 52.40 \% & 0.42 & $1\,626$ \\
    
        & \multirow{1}{*}{Attributes} & 30.64 \% & 43.02 \% & 0.35 & 46.79 \% & 0.36 & 50.94 \% & 0.36 & $1\,325$ \\
    
        & \multirow{1}{*}{Associations} & 35.41 \% & 42.99 \% & 0.39  & 45.18 \% & 0.39 & 46.88 \% & 0.39 & $3\,296$ \\
        
    \midrule
        \multirow{3}{*}{\textbf{Local Context}} & Classes & 27.13 \% & 41.97 \% & 0.33 & 45.88 \% & 0.33 & 50.71 \% & 0.33 & $1\,408$ \\
    
        & \multirow{1}{*}{Attributes} & 24.47 \% & 37.72 \% & 0.30 & 42.31 \% & 0.30 & 46.11 \% & 0.30 & $1\,132$ \\
    
        & \multirow{1}{*}{Associations} & 31.79 \% & 42.69 \% & 0.36 & 44.85 \% & 0.36 & 46.38 \% & 0.37 & $3\,284$ \\

    \bottomrule
    \end{tabular*}
\end{center}

\label{tab:rq1_rq2_results}
\end{table*}

In this section, we present the results of our experiments and answer the research questions.

\subsection{RQ1 -- Metamodel Renaming with Global Context Sampling}
\label{sec:results_rq1}

We start our experiments by determining whether our pre-trained language model is able to recommend meaningful domain concepts (class, attribute, and association identifiers) when considering a global sampling strategy. The upper part of \Table{tab:rq1_rq2_results} shows the top-$1$ accuracy, recall, MRR when varying the size of the recommendation lists. As discussed previously, we consider a recommendation to be accurate when it contains an element that exactly matches the expected identifier (ground-truth). 

The first observation that we can draw is that our approach is able to predict the correct ground-truth (\ie top-$1$) when providing only one recommendation candidate. For the three prediction tests, the top-$1$ score is above $30\%$ and reaches almost $40\%$ for the prediction of classes. This already shows that our model is able to learn domain concepts from our training set and is able to generalize to some extent to unseen metamodels. 

Secondly, we can observe a common increasing trend of the effectiveness of the approach as we increase the number of candidates in the predictions, \ie Recall@k increases when $k$ increases. However, the MRR does not increase like the recall, which is not surprising. Indeed, the MRR@5 of $0.42$, $0.35$ and $0.39$ indicate that the majority of the correct recommendations generally appear among the three first candidates. Consequently, the recommendations that account for the increasing in the Recall@10 and Recall@20 provide the correct candidate at a low rank in the list. Thus, even though increasing the number of candidates increases the recall, it is still able to produce relatively accurate top-$5$ recommendations. This is valuable for modelers to keep the auto-completion list short.

\begin{figure}[!t]
    \centering
    \includegraphics[width=\linewidth]{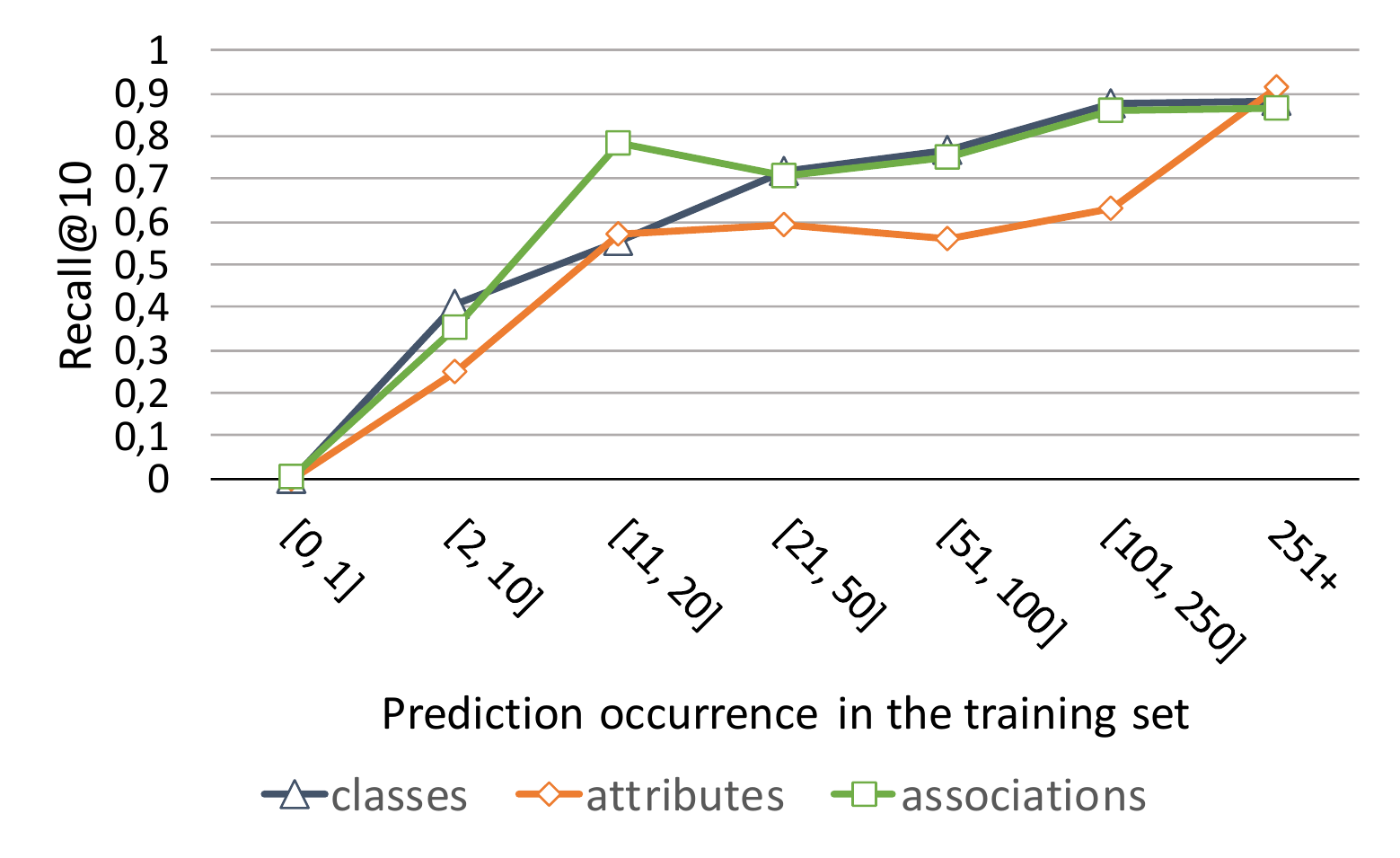}
    \caption{[RQ1] -- Comparison of the Recall@10 obtained w.r.t the occurrences of the recommended concepts in the training set.}
    \label{fig:rq1_occurrence}
\end{figure}

Finally, in \Fig{fig:rq1_occurrence} we analyze the evolution of the Recall@10 for each type of recommendation depending on the occurrences of the recommended concepts in the training set. As we can observe, the model is not able to recommend hapax legomena (see \Table{tab:training_stat}) and unseen concepts. However, the effectiveness keeps increasing as the interval of occurrence increases reaching 80 -- 90\% Recall@10 for the most-common concepts. Thus, uncommon concepts at test time are responsible for a significant drop in the effectiveness of the model as the average Recall@10 reported in \Table{tab:rq1_rq2_results} does not exceed $48\%$. In addition, this level of accuracy is obtained with at least 11 occurrences of the concept, which is already low considering the size of our training set.

Overall, we notice that the model is slightly more accurate when predicting classes. One might think that classes are more difficult to predict than attributes and associations because they embody higher-level concepts than the latter two. This shows that domain concepts of higher level and some of their relationships are probably more recurrent among metamodels than local/low-level concepts. Thus, our pre-trained language model is able to learn meaningful representations of these high-level concepts.
Based on the number of test samples (\ie more than $5\,000$), we can state the following.
\begin{oframed}
\noindent\textbf{Answer to RQ1:} Our approach can recommend relevant domain concepts, that were seen at least twice during the training phase, to rename metamodel elements when it considers a global context to provide predictions.
\end{oframed}
As we will illustrate it in the answer to RQ4, we can improve the quality of the recommendations if we consider partial or semantically equivalent concepts.
\subsection{RQ2 -- Metamodel Renaming with Local Context Sampling}
\label{sec:results_rq2}

\begin{figure}[!t]
    \centering
    \includegraphics[width=\linewidth]{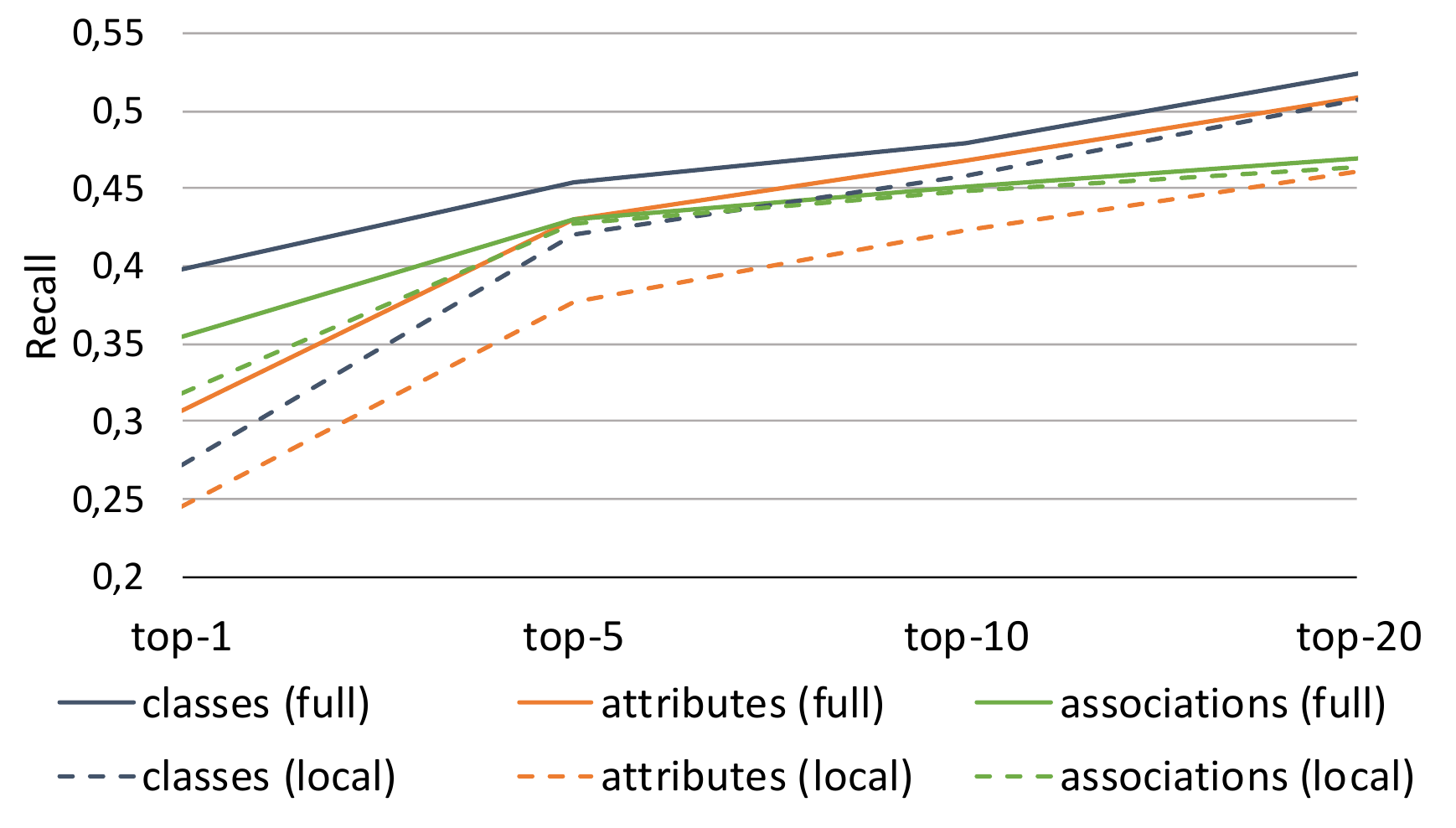}
    \caption{[RQ1 - RQ2] -- Comparison of the level of recalls obtained for each prediction test (full: global context sampling = RQ1; local: local context sampling = RQ2)}
    \label{fig:rq1_rq2_comparison}
\end{figure}

In this second experiment, we compare the performance of both global and local context sampling strategies. \Fig{fig:rq1_rq2_comparison} shows the evolution of the recall obtained for each strategy when the number of recommended candidates increases. 

For the three prediction tests, the local context sampling has a lower recall rate than the global context sampling strategy. These results contradict our conjecture that most of the semantics of an element in a metamodel can be explained locally and that considering other elements might introduce noise to the model. Instead, the results suggest that considering all the elements of the metamodel helps to make the model more accurate. 
\begin{oframed}
\noindent\textbf{Answer to RQ2:} Reducing the context to elements local to the one being predicted does not improve the effectiveness of our apporach.
\end{oframed}
Instead, a global context sampling strategy allows the model to provide more accurate recommendations due to the additional context.
\subsection{RQ3 -- Incremental Construction of a Metamodel}
\label{sec:results_rq3}

We evaluate the capacity of our system to recommend relevant domain concepts to a modeler during the construction of a metamodel. As we progress in the construction of a metamodel, we are able to infer more context to the trained model. Therefore, we evaluate the impact of the context on the effectiveness as it reflects the efficiency of the system through the whole design process. 

\begin{figure*}
    \centering
    \includegraphics[width=.7\textwidth]{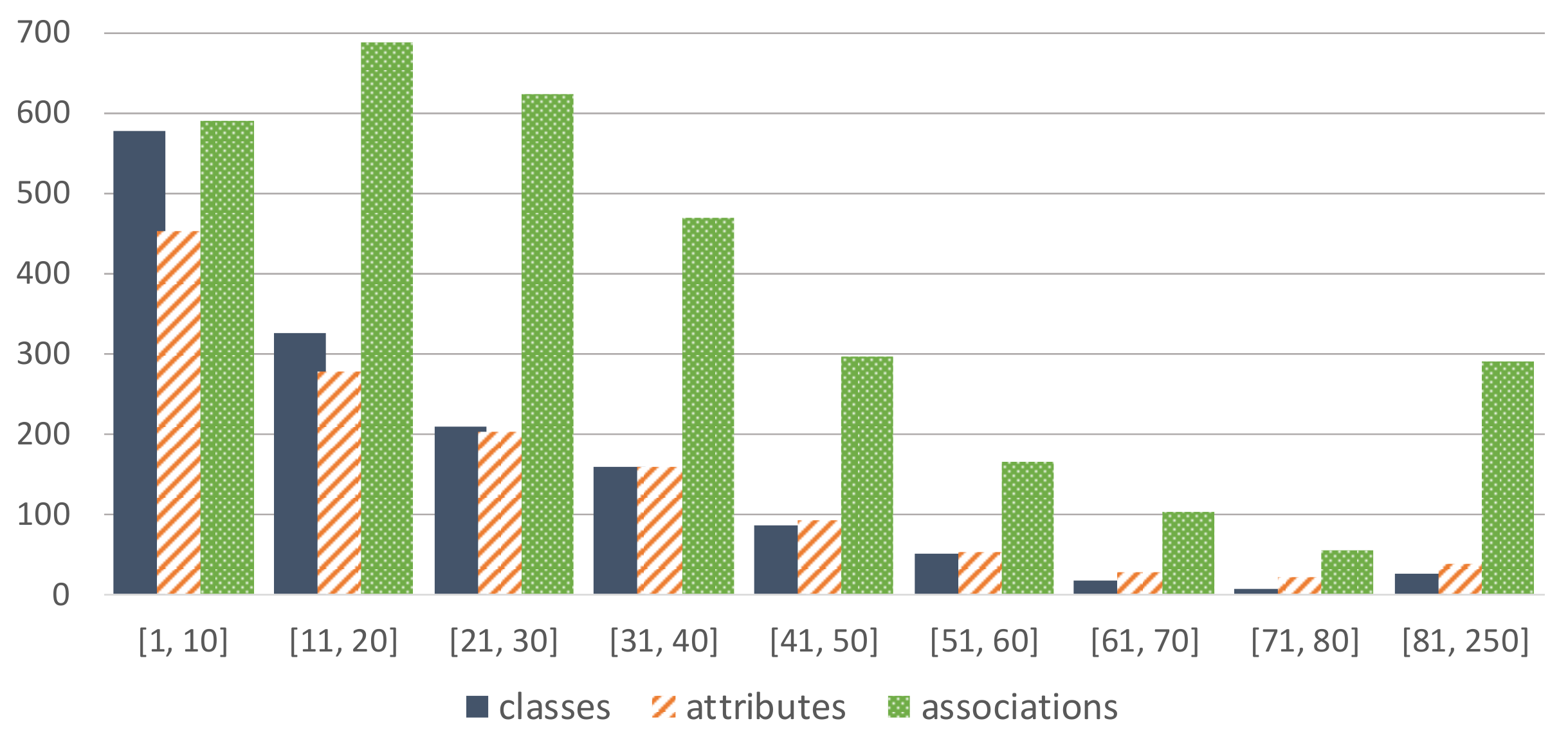}
    \caption{[RQ3] -- Distribution of each type of prediction in the test set according to the number of elements in the available context (\ie metamodel elements that have already been designed).}
    \label{fig:rq3_data}
\end{figure*}

In \Fig{fig:rq3_data}, we report the distribution of the test set in this sampling scenario for different ranges of context, where the context size corresponds to the number of classes, associations, and attributes available (or already designed). On the one hand, it shows that the majority of the test samples are distributed within a context size $\in [1, 40]$. On the other hand, it shows that our test set is relatively diverse in size and allows us to evaluate our approach under fairly broad range of situations. 

\begin{figure}[!t]
    \centering
    \includegraphics[width=1\linewidth]{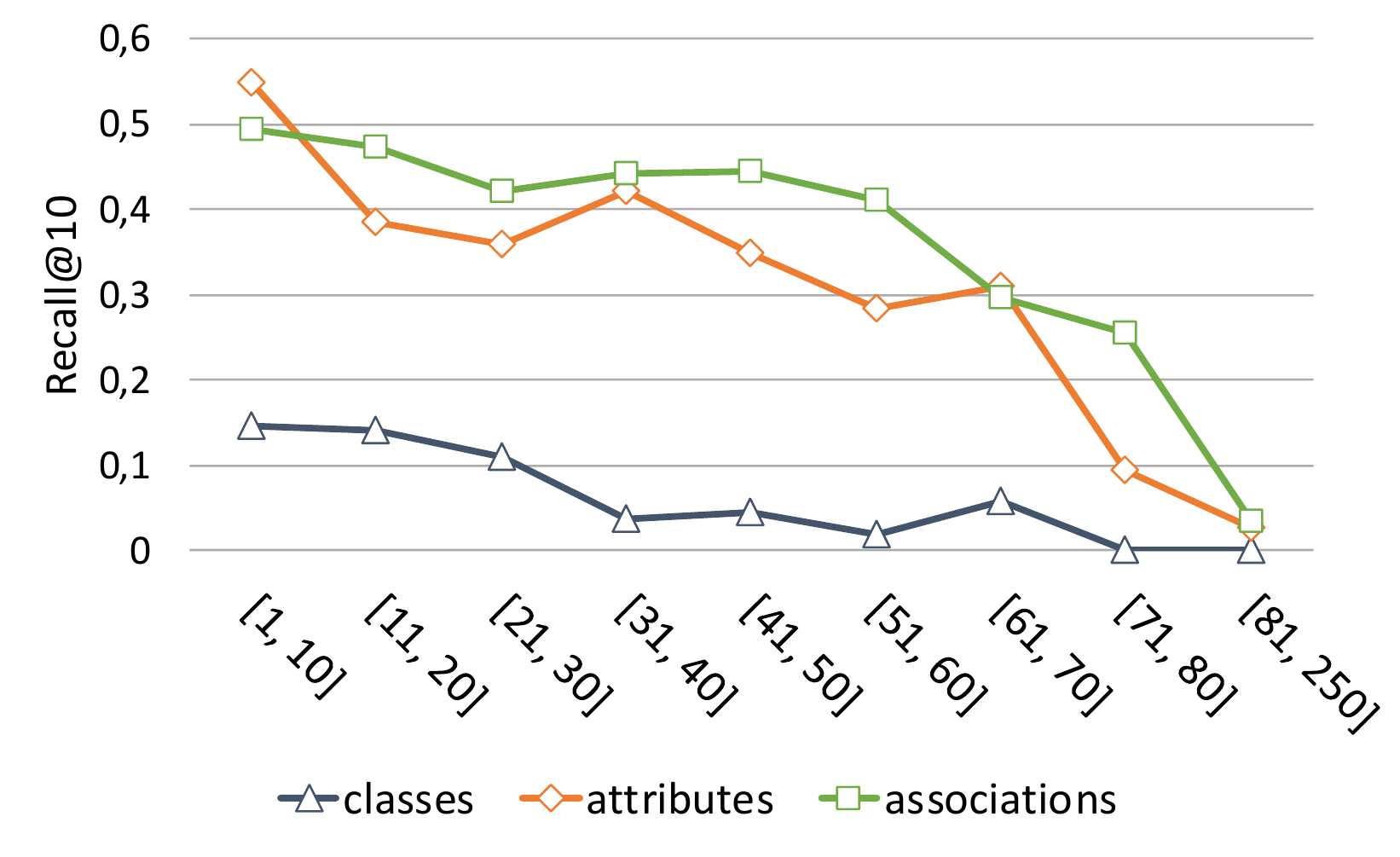}
    \caption{[RQ3] -- Evolution of the Recall@10 with respect to the size of the context}
    \label{fig:rq3_evolution}
\end{figure}

In \Fig{fig:rq3_evolution}, we report the evolution of the Recall@10 according to the size of the context for the three prediction tests. Surprisingly, the effectiveness on classes is substantially lower than that obtained in RQ1 and RQ2. This suggests that classes are in reality more difficult to predict than associations and attributes. Also, providing smaller context \ie from $[1, 10]$ to $[21, 30]$) results in a higher effectiveness than for larger context sizes. This result is not necessarily consistent with those obtained in RQ1, but as the test sampling are totally different (which results in totally different test samples), a thorough comparison of both modeling scenarios is not feasible. However, given the results obtained in RQ1 (\ie $45$\% Recall@5 for the prediction classes), we can hypothetize that the context generated in this modeling scenario is not appropriate or sufficient for the prediction of classes.

In general, we can observe that increasing the context size hinders the results. In the case of attributes and associations, the Recall@10 is higher for context sizes between $[1,50]$ and decreases as the context grows.
However, for the context sizes between $[51,250]$, the number of test samples decreases and may not be representative enough, which can partly explain the drop in the effectiveness.

\begin{oframed}
\textbf{Answer to RQ3:} When incrementally constructing metamodels, we conclude that for low context ranges, the system is relatively efficient to predict associations and attributes, but is less efficient to predict classes.
\end{oframed}
Here again, these results must be taken in the context of an exact match between the predicted and the expected concepts. A more relaxed comparison taking into account partial and semantically-equivalent concepts would result in better predictions as illustrated in \Sect{sec:results_rq4}.
\subsection{RQ4 -- Use Cases}
\label{sec:results_rq4}

In this last experiment, we consider the Petri nets and Java metamodels from our test set. We first evaluate the metamodel renaming scenario with global context sampling which gave us the best performance on the whole test set in RQ1. Then, we evaluate the incremental construction scenario of RQ3 on both metamodels.

\begin{figure*}[!ht]
    \framebox[\textwidth]{
    \begin{minipage}{\textwidth}
    \vspace{5pt}
    \centering
        \begin{minipage}[b]{\textwidth}
            \centering
            \includegraphics[width=.8\textwidth]{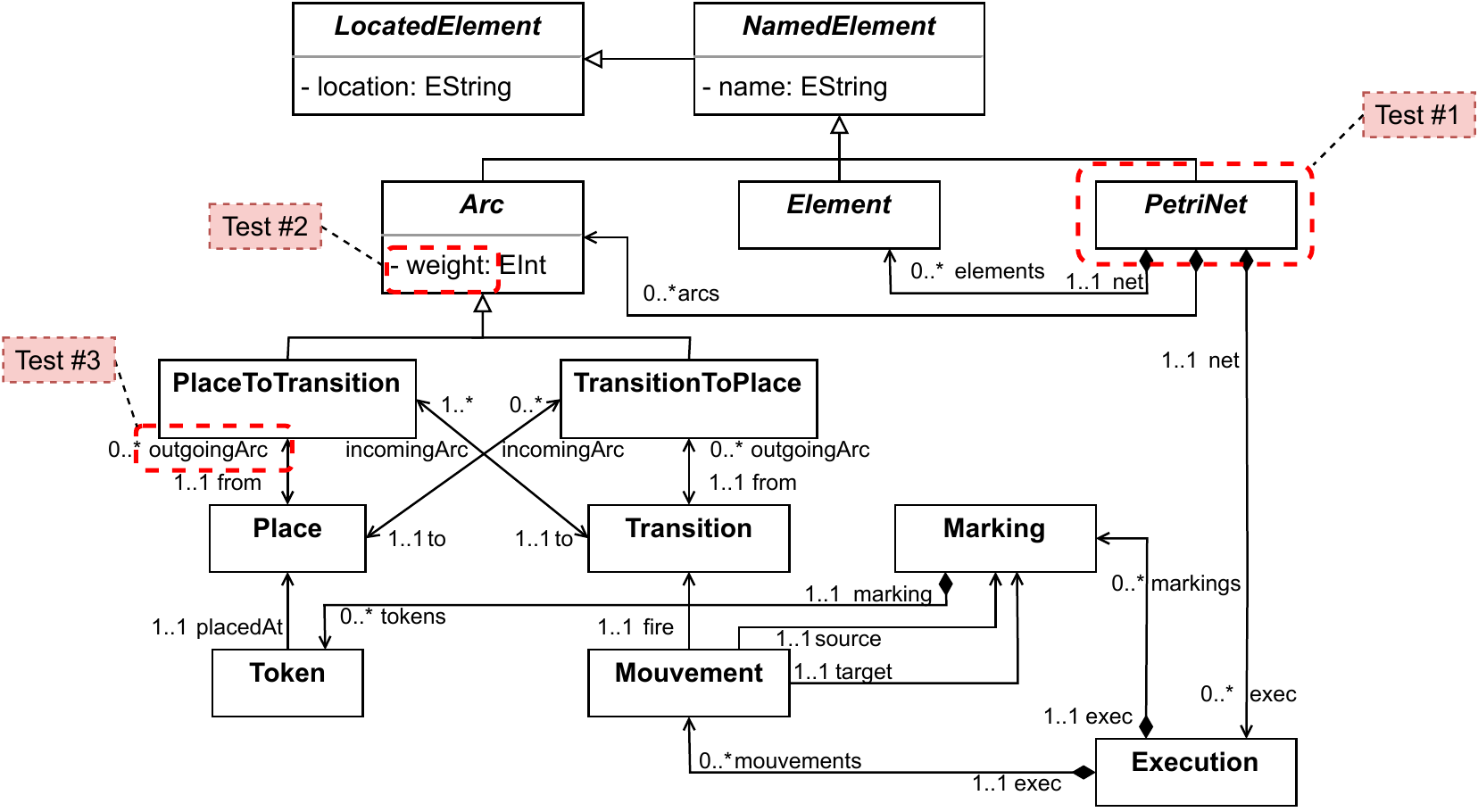}
            \label{}
        \end{minipage}
        \hfill
        \vspace{\floatsep}
        
        \begin{minipage}[b]{\textwidth}
            \centering
            \begin{subtable}[b]{.9\textwidth}
                \renewcommand{\arraystretch}{1.2}
                \setlength{\arrayrulewidth}{.5pt}
                
                \begin{tabular*}{\textwidth}{@{\extracolsep{\fill}}*{4}{l|cccc}}
                \toprule
                    \multicolumn{1}{l}{} & 
                    \multicolumn{1}{c}{\textbf{Test \#1}} &
                    \multicolumn{1}{c}{\textbf{Test \#2}} &
                    \multicolumn{1}{c}{\textbf{Test \#3}} \\
                \midrule
                    \multirow{1}{*}{\textbf{Ground-Truth}} & \multicolumn{1}{c}{PetriNet} & \multicolumn{1}{c}{weight} & \multicolumn{1}{c}{outgoingArc} \\ 
                \midrule
                    \multirow{5}{*}{\textbf{Top-$5$}} & {\color{red} 1. PetriNet} & {\color{red} 1. weight} & {\color{red} 1. outgoingArc} \\
                    & 2. LocatedElement & 2. marking & 2. incomingArc \\
                    & 3. Model & 3. token & 3. linksToPredecessor \\
                    & 4. Petrinet & 4. probability & 4. generatedEvents \\
                    & 5. Graph & 5. kind & 5. incomings \\
                \bottomrule
                \end{tabular*}
            \end{subtable}
        \end{minipage}
    \vspace{5pt}
    \end{minipage}
    }
    \caption{[RQ4] -- Excerpt of the renaming of a classifier, an attribute and an association on the Petri nets use case metamodel using the full context sampling approach (RQ1).}
    \label{fig:rq4_use_case1}
\end{figure*}

\begin{figure*}[!t]
    \framebox[\textwidth]{
    \begin{minipage}{\textwidth}
    \vspace{5pt}
    \centering
        \begin{minipage}[b]{\textwidth}
            \centering
            \includegraphics[width=.9\textwidth]{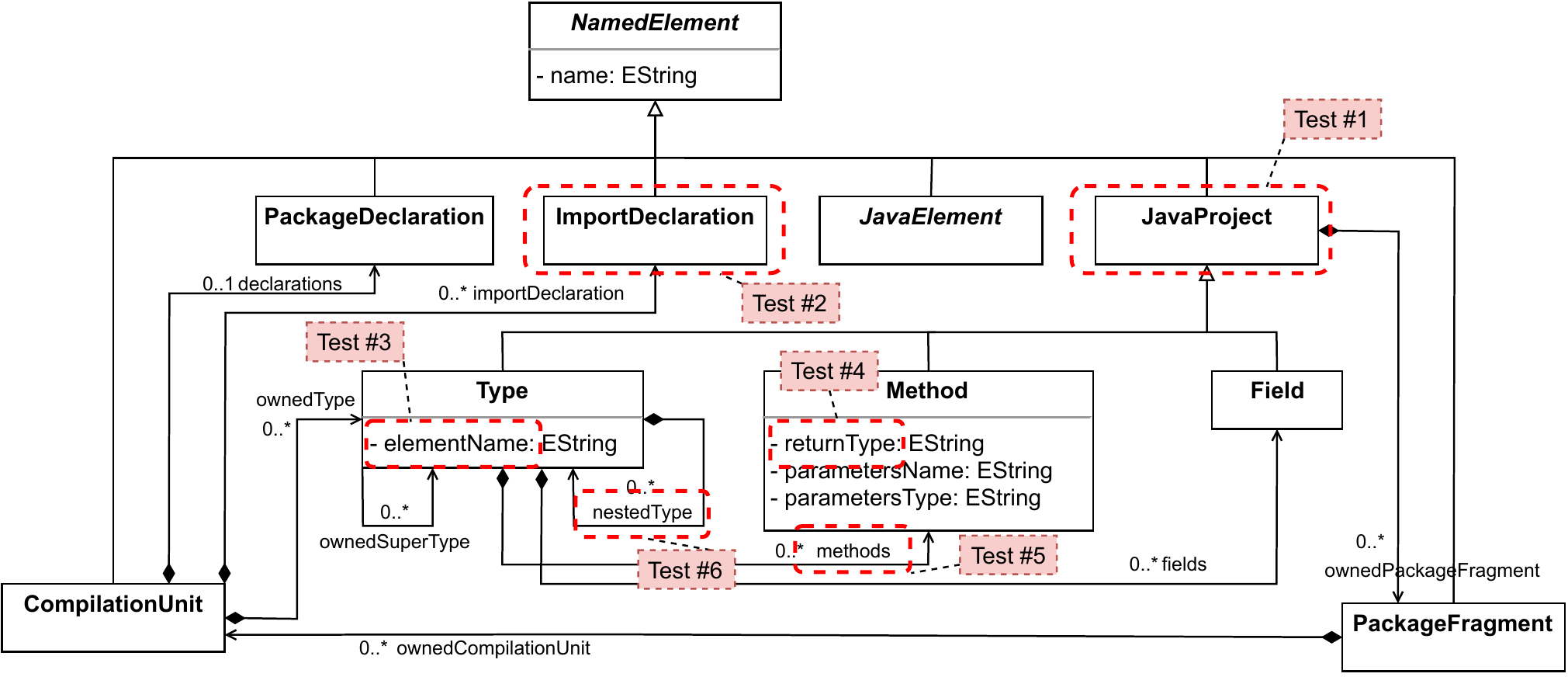}
            \label{}
        \end{minipage}
        \hfill
        \vspace{\floatsep}
        
        \begin{minipage}[b]{\textwidth}
            \centering
            \begin{subtable}[b]{.9\textwidth}
                \renewcommand{\arraystretch}{1.2}
                \setlength{\arrayrulewidth}{.5pt}
                
                \begin{tabular*}{\textwidth}{@{\extracolsep{\fill}}*{4}{l|cccc}}
                \toprule
                    \multicolumn{1}{l}{} & 
                    \multicolumn{1}{c}{\textbf{Test \#1}} &
                    \multicolumn{1}{c}{\textbf{Test \#2}} &
                    \multicolumn{1}{c}{\textbf{Test \#3}} \\
                \midrule
                    \multirow{1}{*}{\textbf{Ground-Truth}} & \multicolumn{1}{c}{JavaProject} & \multicolumn{1}{c}{ImportDeclaration} & \multicolumn{1}{c}{elementName} \\ 
                \midrule
                    \multirow{5}{*}{\textbf{Top-$5$}} & 1. Model & {\color{red}1. ImportDeclaration} & 1. name \\
                    & 2. DomainModel & 2. NamespaceDeclaration & 2. visibility \\
                    & 3. Root & 3. Elements & 3. type \\
                    & 4. Module & 4. rules & 4. returnType \\
                    & 5. Program & 5. ModelingConcept & 5. types \\
                \midrule
                    \multicolumn{1}{l}{} & 
                    \multicolumn{1}{c}{\textbf{Test \#4}} &
                    \multicolumn{1}{c}{\textbf{Test \#5}} &
                    \multicolumn{1}{c}{\textbf{Test \#6}} \\
                \midrule
                    \multirow{1}{*}{\textbf{Ground-Truth}} & \multicolumn{1}{c}{returnType} & \multicolumn{1}{c}{methods} & \multicolumn{1}{c}{nestedType} \\ 
                \midrule
                    \multirow{5}{*}{\textbf{Top-$5$}} & 1. name & {\color{red}1. methods} & 1. type \\
                    & 2. methodName & 2. method & 2. returnType \\
                    & 3. parameterType & 3. Methods & 3. types \\
                    & {\color{red} 4. returnType} & 4. definitions & 4. superType \\
                    & 5. elementName & 5. parameters & 5. parameters \\
                \bottomrule
                \end{tabular*}
            \end{subtable}
        \end{minipage}
    \vspace{5pt}
    \end{minipage}
    }
    \caption{[RQ4] -- Excerpt of the renaming of two classifiers, two attributes and two associations on the Java use case metamodel using the full context sampling approach (RQ1).}
    \label{fig:rq4_use_case2}
\end{figure*}

In \Fig{fig:rq4_use_case1} and \ref{fig:rq4_use_case2}, we report several top-$5$ recommendations provided by our model using the sampling of RQ1 for both test metamodels. The dashed boxes represent the mask used for the tests.
At first, we can observe that the recommendations for the Petri nets metamodel are very relevant and perfect in the three given examples. Moreover, the selected test samples are representative of the overall results obtained on this metamodel. In fact, we obtain $100\%$ Recall@5 for both classes and attributes, and $82\%$ Recall@5 for the associations. Besides, the top-$5$ lists show that the model is able to recommend domain concepts that are related to Petri nets. This provides the modeler with relevant concepts applicable in the current modeling activity. 

For the Java metamodel (\Fig{fig:rq4_use_case2}), the results show a decrease in effectiveness. We obtain $60\%$, $33\%$ and $22\%$ Recall@5 for the classes, attributes and associations, respectively. The expected recommendation is not always in the top-5 list. Nevertheless, we note that all recommendations are concepts close to the application domain related to Java. For instance, in Test \#1, the model is able to identify that the \Code{JavaProject} class represents a high-level concept that could be matched with the concepts of \Code{Program} or \Code{Module}. Overall, the recommendations appear to be coherent with the application domain and still provide useful domain concepts to the modeler. 
\begin{figure}[!t]
    \centering
    \includegraphics[width=1\linewidth]{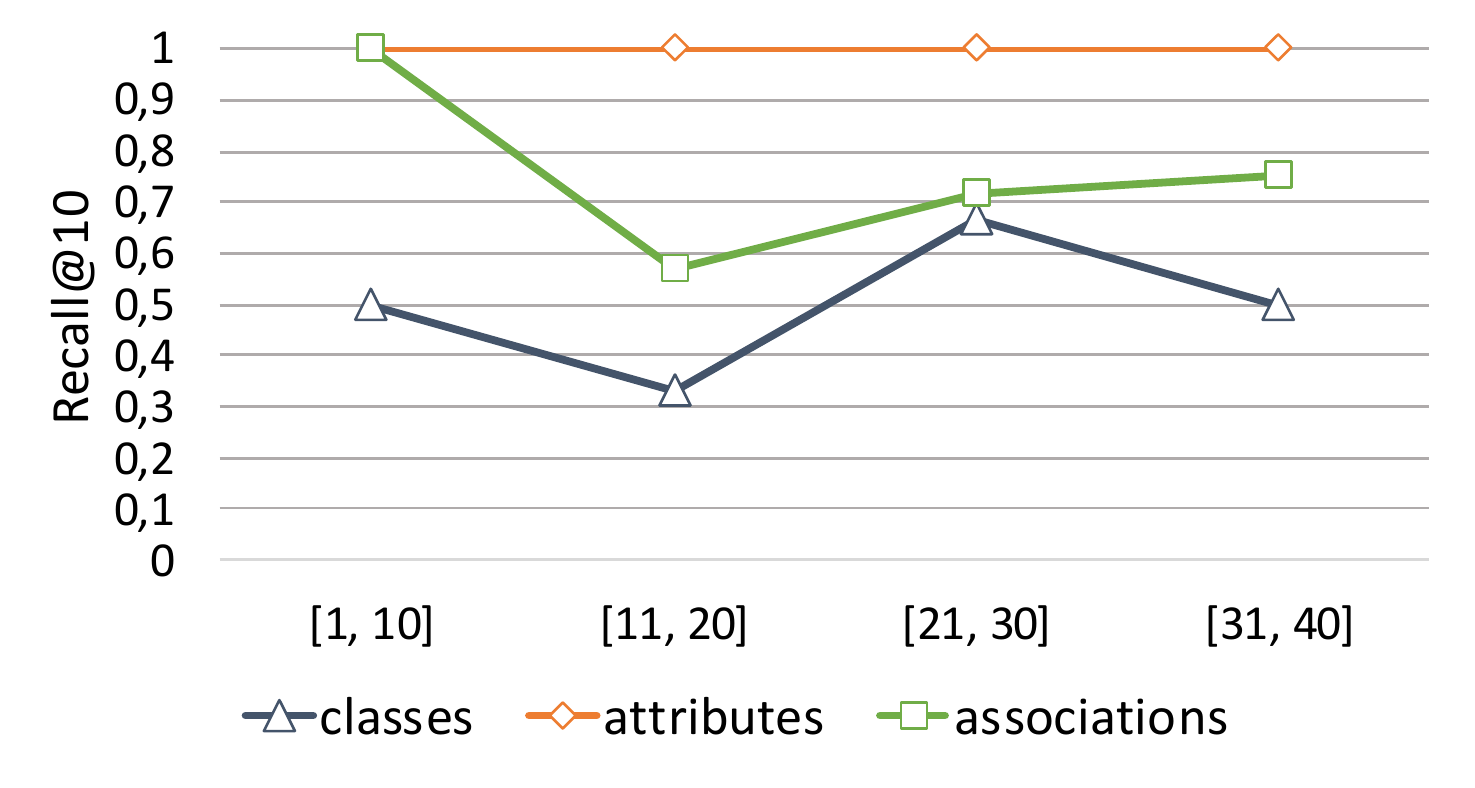}
    \caption{[RQ4] -- Evolution of the Recall@10 with respect to the size of the context for the Petri nets use case in the modeling scenario of RQ3.}
    \label{fig:rq4_Petri nets_construction}
\end{figure}

Considering the modeling scenario of RQ3, we report in \Fig{fig:rq4_Petri nets_construction} the evolution of the Recall@10 with respect to the context size for the Petri nets metamodel. It is a medium-sized metamodel and the trends in the results corroborate with those obtained in RQ3, except that the Recall@10 is much higher for this specific metamodel. This can be explained by the fact that the Petri nets metamodel was defined using relatively general concepts (\eg \Code{NamedElement}, \Code{Place}, \Code{Transition}, \Code{Execution}) that are frequently used in the design of other metamodels present in our training set. However, the results for the Java metamodel in this scenario are much lower.
We envision that in a modeling activity similar to the one of RQ3, providing the model with additional contextual information could help improving the results. In fact, as shown in \Fig{fig:rq4_use_case2}, when we provide the model with a global context, it is able to provide relevant recommendations. Therefore, we believe that in an incremental construction scenario, this lack of context could be bridged by taking into account data source other than metamodels, \eg requirement specifications.

\begin{oframed}
\textbf{Answer to RQ4:} Our model performs well in a renaming scenario for two metamodels which use both general and specific domain concepts. For an incremental construction process, the model provides meaningful domain concepts in the first use case where the modeler manipulates general concepts, but lacks in effectiveness in the second use case where the concepts are more domain-specific.
\end{oframed}

\subsection{Threats to Validity}
\label{sec:results_threats}

We identified some threats to the validity of our evaluation and attempted to address them during its design. A first external threat to the validity relates to the chosen representation of the metamodels and how the trees are organized. We chose to transform the metamodels into trees by following the ``natural order'' in which the elements of the metamodels appear in the corresponding Ecore serializations. We did not compare this approach with other tree structures that could eventually result in a better effectiveness of the system. Nevertheless, we enabled the approach to be general enough to use different structures of the trees with slight adaptations. 

The design of our modeling scenarios results in the main threat to construct validity. We attempted to articulate our research questions to match realistically simulated modeling scenarios. In the first one (\ie RQ1 -- RQ2), we believe that the two sampling strategies are reasonable since they are applicable in a real situation where a metamodel is already designed. In the second one (RQ3), we mitigated the threat as best as we could by implementing a sampling algorithm to get as close as possible to a real-world situation. Since there is an infinite number of possibilities to design a metamodel (due to human choices or modeling constraints), we believe that the proposed simulated scenarios are reasonable to properly evaluate our approach.

Next, an internal threat to the validity that could prevent from strictly replicating of our experiments involves the training of our pre-trained language model and the choice of the hyperparameters. The initialization of neural networks parameters (\ie weights) is random and two models trained with the same hyperparameters could lead to slight variations in their accuracy. Also, the choice of hyperparameters is crucial to optimize the model to improve its generalization to unseen data. To mitigate this threat, we tuned the hyperparameters of the RoBERTa architecture using ranges of values widely used in the literature. We reported the optimal model configurations in Appendix \ref{sec:model_hp}.

Finally, another important aspect that we considered is the representativeness of our data. We trained our model on $10\,000$ metamodels crawled from open-source repositories with a rather broad range of sizes and various application domains. Regarding the evaluation, we use a completely separate dataset from the one used for training, with metamodels covering a variety of application domains and of various sizes.
\subsection{Discussion}
\label{sec:results_discussion}

We investigated the idea of learning domain concepts from metamodels covering a various range of application domains using a state-of-the-art pre-trained language model architecture. We regard this work as an initial investigation that presents some limitations but also that our experiments highlight interesting initial results in the context of assisting a modeler in a modeling activity. We identified a number of lessons learned as well as opportunities to improve our approach in future work.

    Regarding the results of our experiments, we showed that our model is less efficient on domain-specific metamodels (\eg Java metamodel), which causes a drop in the effectiveness. These results could also be explained by the fact that some concepts do not appear often in the training data, which may result in the inability of the model to predict them. We envision that this lack of efficiency of our approach on these types of metamodels could be overcome in two ways. One option is to consider knowledge data (\eg requirements, modeling environment data, ontologies) in our framework, similarly to related work~\cite{agt2018automated,burgueno2020nlp}. Alternatively, we could recommend generic terms and let the modeler decide of the best combination of terms to define her specific concepts.
    
    Through the modeling scenario of RQ3, we simulated a real-world incremental construction of a metamodel and highlighted mitigated results. We plan to further investigate this particular scenario in future work by conducting a case study on the usability of our approach in a real modeling environment, with real users. Additionally, we are convinced that considering the user in the loop is important in this process and could lead to significant improvements of our approach. We foresee that reinforcement learning or search-based techniques with user interactions could be applicable in this scenario.
    
    To evaluate the accuracy of our predictions, we used an exact match strategy. In the real world, modelers can use different names for the same concepts, especially when these names are combinations of multiple terms. In the future, we plan to use semantic similarity metrics based on, for example WordNet\footnote{
      \url{https://wordnet.princeton.edu/}
    }, to compare the predicted and the expected concepts. We believe that this strategy will allow a better evaluation of our approach. 
    
    In our proposed representation of metamodels as trees, we purposely did not consider all the elements of the metamodels as we discussed that some of them may not be relevant considering our modeling tasks. Nevertheless, we argue that including more elements in our representations (\eg generalizations, enumerations, cardinalities) could be valuable in particular in the context of modeling scenarios similar to the ones tackled in this work. As discussed previously, we designed our approach to be independent from the representation of the metamodels. Therefore, we plan to explore alternative representations of the metamodels in future work and compare them on real modeling scenarios without necessarily having to alter our approach. 
    
    In this work, we used a RoBERTa pre-trained language model architecture to learn metamodels domain concepts. In all of our experiments, we evaluated the trained model without fine-tuning procedure. In spite of that, the reported results are still promising and performing a fine-tuning step beforehand the evaluation could lead to an increasing in the effectiveness of the approach. Additionally, the pre-trained language model that we released can be fine-tuned for a broad range of other tasks. 
   
    On the learning model used in this work, we worked with a state-of-the-art bidirectional language model that deals with sequential data as input. As discussed in \Sect{sec:learning_domain_concepts}, we had to flatten the trees into sequences of tokens to train the model. Nevertheless, we concede that metamodels are not sequential and that one may think that a language model may not be suitable given the problem addressed in this paper. However, through our experiments we have shown that such language model is still able to learn meaningful domain concepts. Nevertheless, other model architectures, such as graph neural networks based on graph-structured representation of the input, might be a relevant alternative to better learn from the structural aspect of the metamodels~\cite{scarselli2008graph,kipf2016semi}.
    
    We proposed a fully data-driven approach to design an intelligent modeling assistant. As opposed to source code, the amount of available data in MDE is significantly lower. We believe that if we had more datasets of metamodels at our disposal, we would be able to significantly improve the effectiveness of our approach. However, there is no indication that we will be able to gather more MDE data in a near future and in the same order of magnitude as that available for source code. Consequently, we conjecture that to tackle this lack of data, we need to design hybrid approaches as mentioned previously in this section.


\section{Related work}
\label{sec:related_work}
Intelligent modeling assistants are similar to recommendation systems for software engineering~\cite{robillard2009recommendation, mussbacher2020opportunities} and aim at supporting a modeler through a modeling activity by providing the modeler with relevant recommendations to the modeling context. We present a summary of the existing related work in modeling assistants for MDE with a focus on intelligent assistants to the complete of models and metamodels.

The early works on model completion focused on the design of systems based on rules and logic programming~\cite{Sen2007DSM_completion,Sen2007Partial_completion}. The authors focused on the completion of partial models by deriving constraint logic programs from models using model transformations. The defined approaches are rather complex as they involve many model transformations and a Prolog engine. In comparison, our approach does not require to define any constraints and model transformations manually to provide relevant recommendations as it is fully data-driven. Rabbi\etal{}~\cite{Diagrammatic_completion} designed an approach to rewrite partial models that ensures their conformance to their metamodels. They defined the completion rules using model transformations. In this work, we propose an approach that leverages the structural and lexical properties of metamodels in a large corpus by means of a language model that does not require any pre-defined rule. 

Kuschke\etal{}~\cite{kuschke2013recommending} defined an approach that captures editing operations during modeling activities to recommend further modeling activities. Their approach is based on a catalog of common UML modeling activities to produce the recommendations. They extended their work by allowing more control from the modeler~\cite{kuschke2014pattern}. In contrast, our system is independent from any modeling environment as it does not rely on data extracted from the latter. Moreover, we defined a general approach that is not restricted to the completion of specific cases that need to be defined manually. 

Most recent works have focused on gathering knowledge data by extracting meaningful information from textual or structured data (\eg requirements, application domain, models/metamodels). Elkamel \etal{}~\cite{elkamel_uml_recommender_2016} designed a system to recommend UML classes during a modeling activity. They defined a clustering-based approach whose objective is to find similar UML classes to the current modeling context. Although the approach is interesting, it lacks evidence whether it can scale to large datasets. Alternatively, Stephan~\cite{stephan2019towards} proposed a similarity-based approach to guide the modeling process by considering elements from clone models to the partial model being designed as possible recommendations. Yet, as we discussed in \Sect{sec:results_discussion}, we envision that clustering-based could be an interesting direction to develop an alternative approach based on semantic similarity for the completion of metamodels. 

Agt-Rickauer\etal{}~\cite{agt2018automated} proposed a domain modeling recommender system based on knowledge data of domain-specific terms and their relationships. The system combines various sources of data containing ontologies, concepts, and terms that allow the completion of named elements in a model. In another work, Agt-Rickauer\etal{}~\cite{agt2018domore} used a frequency-based approach to recommend concepts in domain models. Their approach requires a large amount of background knowledge data in order to generalize to a certain range of domains.
In contrast, our approach does not rely on ontologies or vocabularies that are difficult to obtain in an exhaustive way. Instead, we learn implicitly the relationships between the domain concepts instantiated in thousands of metamodels using a pre-trained language model which makes our work independent from any application domain. 

More recently, Di Rocco \etal{}~\cite{models21-gnn-recommender} proposed a GNN-based approach to recommend classes and class members in models. Similarly to our work, they found that classes are difficult to recommend due to the broad range of concepts used by modelers. Burgueno\etal{}~\cite{burgueno2020nlp} designed a NLP-based approach to assist a modeler based on textual knowledge and general knowledge data. They also enabled their assistant to take into consideration the user feedback by monitoring its interaction with the assistant. However, the proposed approach requires to extract relevant sources of general knowledge to gap the lack of contextual knowledge; as it may not always be accurate when a modeling activity requires knowledge about a specific application domain. In our approach, we do not require to gather contextual data about the metamodel domain to provide relevant recommendations. Instead, we learn metamodel domain concepts using a pre-trained language model architecture. Nevertheless, we argue that our approach presents some limitations and extending it to allow processing general knowledge data or contextual data (\eg requirements) would undoubtedly improve its effectiveness. Additionally, integrating the user feedback similarly to Burgueno\etal{}'s work is an interesting opportunity for future work as discussed in \Sect{sec:results_discussion}.
Finally, Mussbacher\etal{}~\cite{mussbacher2020towards} designed a framework to evaluate intelligent modeling assistants on their ability to learn from data and their usefulness to an actual modeler. As a future work, we plan on assessing our intelligent modeling assistant using their assessment grid.

\section{Conclusion}
\label{sec:conclusion}
In this paper, we presented a novel approach to learn metamodel domain concepts using a RoBERTa pre-trained language model architecture based on tree representations of the metamodels. With this model, we built a recommender system that is able to assist a modeler during a modeling activity by recommending meaningful domain concepts relevant to the modeling context. 

As our first contribution, we presented an approach to transform a metamodel into a tree structure that allows to capture both structural and the lexical information contained within the metamodel. Then, we have introduced our overall framework to learn domain concepts based on the aforementioned representations using a state-of-the-art pre-trained language model architecture. 

We evaluated our approach on three simulated modeling scenarios and demonstrated the applicability of our approach in these scenarios. The experiments highlight promising results and suggest that our pre-trained language model can learn domain concepts from a dataset of independent metamodels without requiring additional data about the application domains. The model and data are publicly available (see Appendix \ref{sec:replication_package}).

Although we provide compelling evidence that it is possible to recommend meaningful concepts during the modeling activities, there is still room for improvement to better assist modelers. We intend to develop more complex intelligent modeling assistant that would consider the human in the loop and her feedback during a modeling activity. We will explore other model architectures such as graph neural networks and compare their effectiveness on real modeling tasks. Finally, we plan on implementing a hybrid approach that considers knowledge data, such as requirements, to improve the effectiveness of our approach on domain-specific modeling activities.


\bibliographystyle{spmpsci}
\bibliography{bibliography}

\appendix
\section{Replication Package}
\label{sec:replication_package}
We make our code, datasets and models publicly available to ease the replication of our experiments and to help researchers that are interested in extending our work: \\ \url{https://github.com/martin-wey/metamodel-concepts-bert}

\noindent The data and models are available on Zenodo: \\
\url{https://doi.org/10.5281/zenodo.5579980}

\section{Model Hyperparameters}
\label{sec:model_hp}

\begin{table}[!h]
\caption{Hyperparameters (HP) used for the training of our model.}
    \centering
    \renewcommand{\arraystretch}{1.2}
    \setlength{\arrayrulewidth}{.5pt}
    \begin{tabular}{lc} \toprule
        HP & Value \\ \midrule
        Number of layers & 12 \\
        Hidden size & 768 \\
        FFN inner hidden size & 3072 \\
        Attention heads & 12 \\
        Dropout & 0.1 \\
        Attention dropout & 0.1 \\
        Hidden activation & gelu \\
        Positional embedding & absolute \\
        Batch size & 32 \\
        Max epochs & 100 \\
        \midrule
        Vocabulary & byte-level BPE \\
        Voocabulary size & 30.000 \\
        Vocabulary cut-off & 2 \\
        \bottomrule
    \end{tabular}
    \label{tab:training_set}
\end{table}

\end{document}